\documentclass{article}

\usepackage{graphicx}   
\usepackage{amsmath}   
\usepackage{cite}       
\usepackage[margin=1in]{geometry} 
\usepackage{authblk}    
\usepackage{subcaption} 
\usepackage{url}        
\usepackage{siunitx}
\usepackage{hyperref}
\usepackage{amssymb}
\usepackage{booktabs}

\title{Study of energy deposition in the coolant of LFR}
\author[1]{Maria Susini}
\author[2]{Sacha Barré}
\author[2]{Daniele Tomatis\thanks{Corresponding author: daniele.tomatis@newcleo.com}}
\author[1]{Stefano Argirò}

\affil[1]{University of Torino, Department of Physics, Torino, Italy}
\affil[2]{newcleo Srl, Via Giuseppe Galliano 27, 10129 Torino, Italy}
\date{\today}

\begin{document}
\maketitle

\begin{abstract}
    The determination of the fraction of energy deposited in the coolant is required for the setup of accurate thermal-hydraulic calculations in reactor core analysis. This study focuses on assessing this fraction and analysing the neutronic and photonic processes contributing to energy deposition in Lead-cooled Fast Reactors (LFRs). Using OpenMC, coupled neutron-photon transport calculations were performed within a fuel pin cell geometry, representative of the one under development at \textsl{new}cleo. Additionally, the implementation of lattice geometry was tested to gauge the impact of reflective boundary conditions on computational efficiency. In the context of a surface-based algorithm, the pin geometry has proven to be computationally more cost-effective. The fraction of energy deposited in the LFR coolant was evaluated at $\sim5.6$\%, surpassing that of pressurised water Reactors ($\lessapprox 3\%$), with photon interactions emerging as the principal contributor. The influence of bremsstrahlung radiation was also considered, revealing minor impact compared to other photonic processes. Given the continuous exploration of various core designs and the expectation of diverse operational conditions, a parametric analysis was undertaken by varying the coolant temperature and pitch values. Temperature changes did not significantly affect the results, while modifying the pitch induced a rise in the fraction of deposited energy in lead, highlighting its dependence on the coolant mass. This mass effect was explored in various fuel assembly designs based on the ALFRED benchmark and on the typical assembly model proposed by \textsl{new}cleo, leading to a correlation function for the prediction of coolant heating in realistic assemblies from simple pin cell calculations.
\end{abstract}

\section{Introduction}
Nuclear fission is expected to play a significant role in the future energy mix thanks to its efficient utilisation of primary resources, carbon neutrality and low environmental impact~\cite{IEA2022}. The evolution of nuclear energy technology towards its next stage involves exploring novel designs~\cite{PIORO2016, GIF2014}, with Lead-cooled Fast Reactors (LFRs) standing out for their enhanced safety features and effective energy production capabilities~\cite{SMITH2016}.
The primary energy released in LFRs results from fission processes induced by fast neutrons within the fuel. Approximately 85\% of this energy is attributed to the kinetic energy of charged fission products, which is predominantly deposited within the fuel through Coulomb interactions. Instead neutral particles, namely neutrons and photons emitted from fission or decay events, can escape the fuel and interact with other materials of the reactor. This collisional transport allows particles to deposit energy away from their emission point, affecting the cladding, coolant, and other structural elements. Given that neutrons and photons contribute about $10$\% of the total energy release, their impact is significant and warrants careful consideration in simulations. In particular, the knowledge of the amount of energy deposited in the coolant is important for general reactor safety and design of the reactor core, since its underestimation can limit operations and economic feasibility, while its overestimation can compromise safety. This aspect was previously studied for a pressurised water reactor (PWR) fuel pin cell where water density and the concentration of boron diluted in the water moderator were varied~\cite{KINAST2021}. In that study, the fraction of energy deposited in water was evaluated at less than $3$\%. A comparable study is lacking for LFRs, where a different behaviour of lead under radiation is expected compared to water.
The goal of this study was to first analyse the physical mechanisms involved in energy deposition in lead by considering various steady-state conditions. This led to the investigation of the performance difference between two geometrical configurations, namely a fuel pin cell and a lattice of fuel pins, both representative of the design under development at \textsl{new}cleo. Additionally, it involved a comparative analysis of the two ways in which OpenMC handles secondary charged particles on heating calculations.  Armed with a better understanding of the mechanisms of energy deposition in lead, the second objective was to devise of relation capable of predicting the fraction of energy deposited in lead in two distinct fuel assemblies based on the ALFRED benchmark~\cite{ALEMBERTI2014} and on the \textsl{new}cleo design.

The present analysis is based on coupled neutron-photon calculations performed at steady-state conditions and with fresh fuel. The contributions from activated products and spent fuel were neglected. The eigenvalue calculations were carried out using OpenMC~\cite{ROMANO2015} v.0.13.4. and the ENDF/B-VIII.0~\cite{BROWN20181} nuclear data library. OpenMC is an open-source Monte Carlo neutron and photon transport simulation code that uses surface-tracking. It originates from the efforts of MIT's Computational Reactor Physics Group and has since garnered contributions to its development from numerous universities, laboratories and organisations.

\section{Physical processes in energy deposition}\label{sec:theory}
Fission stands as the fundamental reaction in reactor operation, representing the primary source of thermal energy release. This energy output depends on both the incident neutron energy and the target nuclide. For instance, when $^{239}$Pu undergoes fission upon interaction with a 1~MeV neutron, $\sim205$~MeV are released~\footnote{Fission energy release for incident-neutron data obtained from the ENDF/BVIII.0~\cite{BROWN20181} library (MF=1, MT=458)}. The released energy is distributed among the particles generated by fission. The predominant portion of released energy is associated with the kinetic energy of fission products, accounting for approximately 175~MeV. Typically emitted in an excited state, these products transition to the ground state through the emission of delayed neutrons, photons and beta particles. While the energy from delayed neutrons can be disregarded due to the low precursor yield fraction (10$^{-3}$), the same cannot be said for delayed photons (5.1~MeV), beta particles (5.2~MeV), and neutrinos (7.0~MeV). Prompt neutrons and photons are produced by fission, carrying kinetic energies of 6.4~MeV and 7.6~MeV, respectively. From this example, it is evident that over 90\% of the released energy is attributed to the production of prompt particles, while the remaining portion is released gradually from the decaying fission products. Among these contributions, only the energy associated with neutrinos is considered lost due to their low interaction cross-section with matter~\cite{FORMAGGIO2012}. Similar orders of magnitude are observed for other fissile nuclides.

The kinetic energy of charged particles is deposited locally in the fuel through electromagnetic interactions, while neutral particles can escape and transfer a portion of their energy elsewhere, at each collision.

In fast reactors, the process of energy deposition differs from thermal ones. The energy lost by neutrons in elastic scattering events varies according to the mass number of the target nucleus; for instance, while hydrogen causes on average 50\% energy loss, $^{208}$Pb results in merely a 1\% loss. Furthermore, lead exhibits low neutron absorption cross section. This dual feature of limited energy loss during scattering and low neutron absorption makes it an excellent coolant choice for fast reactors. Besides the mentioned processes, there is a minor occurrence of inelastic scattering events, which only gain significance at energies above 2.5~MeV.

On the contrary, photons primarily transfer their energy to heavier atoms, resulting in rapid energy loss in lead and significant coolant opacity. As fast reactors operate at high energies, pair production is expected to play a role, albeit minor, in energy deposition, alongside coherent (Rayleigh) scattering, incoherent (Compton) scattering and through the photoelectric effect. Primary interactions also trigger secondary processes which induce the generation of new photons through atomic relaxation, electron-positron annihilation and bremsstrahlung radiation. All these interactions are modelled in OpenMC to accurately replicate photon physics within the reactor and precisely calculate the distribution of energy deposition through coupled neutron-photon transport. The implementation of these calculations in OpenMC is discussed in the section below.

\section{Energy deposition calculations in OpenMC}\label{ssec:openmc}
OpenMC offers two simulation modes. In neutron-only calculations, photons are not transported, whereas they are in the coupled neutron-photon calculations. Neutron transport in the latter serves to establish the source distribution used in the subsequent photon calculation. OpenMC does not account for photon-induced neutron production reactions, which can be neglected for many reactor types. This one-way coupling simplifies problem handling, streamlining computations. It is important to note that these two modes also involve different techniques for calculating energy deposition. Neutron-induced heating is estimated using KERMA~\cite{osti_1338791} (Kinetic Energy Released in MAterials) coefficients modified to include all relevant contributions from fission events. These coefficients have units of eV$\times$barn and can be employed in tally calculations of energy deposition.
Photon heating differs in simulation modes: in neutron-only calculations, photons deposit their energy locally, while in coupled simulations, their energy is transported and deposited through primary interactions within the system. In the latter case, heating is computed as the difference between their pre-collision energy and the combined energy of post-collision photons and secondary particles~\cite{ROMANO2020}.
OpenMC implements two corrections to accurately evaluate energy deposition, as described in~\cite{ROMANO2020}. The first one concerns delayed photons, which are not simulated in a Monte Carlo calculation. Assuming the same energy spectrum for both delayed and prompt photons, in a coupled calculation, the delayed contribution is accounted for by multiplying the yield of prompt photons by the following factor
    \begin{equation}
        f = \frac{E_{\gamma,p}+E_{\gamma,d}}{E_{\gamma,p}}
    \end{equation}
where $E_{\gamma,p}$ and $E_{\gamma,d}$ are the prompt and delayed photons energies, respectively, derived from the ENDF MF=1, MT=458 data. A similar correction is implemented in Serpent~\cite{LEPPANEN2015142}. The second correction aims to solve the imbalance between energy release and deposition caused by the biasing of the fission source by 1/k$_{eff}$ in k-eigenvalues calculations~\cite{jne2020020}. To address this imbalance, the non-fission energy deposition is normalised by the latest estimate of k$_{eff}$ 
    \begin{equation}
        \Tilde{k}(E)=\big(k(E)-k_f(E)\big)k_{eff}+k_f(E),
    \end{equation}
where $k$ and $\Tilde{k}$ are the total KERMA factors before and after correction, and $k_f$ is the fission KERMA. Furthermore, the weight of photons produced from non-fission reaction is scaled by k$_{eff}$.

Energy deposition by electrons and positrons presents unique aspects within OpenMC. Although they are not transported, some of their energy can be used to produce bremsstrahlung photons, which can travel away from the interaction sites. This photon generation is simulated using the thick target bremsstrahlung approximation, where it is assumed that a charged particle's entire energy loss occurs within a single, uniform material region. The number and energy of emitted photons are sampled from probability distribution tables derived from the evaluations of the total stopping power and bremsstrahlung cross-sections specific to the material. Bremsstrahlung photons are emitted with energy ranging from zero to that of the electron or positron, and emerge from the same location and direction as the charged particle. Further information about the implementation of bremsstrahlung and other secondary processes in OpenMC can be found in~\cite{LUND2018}. Bremsstrahlung is the only process users can deactivate by setting the \texttt{electron\_treatment} element to \texttt{led} for local deposition of electrons/positrons energy. The \texttt{ttb} option allows generating bremsstrahlung photons.

Different filters can be applied to define the region of phase space where energy deposition should be scored. One of the filters OpenMC implements isolates contributions according to the particle inducing the reaction. The available particles that can be used as bins for this filter are neutrons, photons, electrons and positrons. The first bin includes contributions from the kinetic energy of fission products, neutrons and beta particles. The subsequent three bins are related to photon heating. Specifically, the electron and positron bins account for locally deposited energy by these charged particles, while the energy used to produce bremsstrahlung photons flows into the photon bin. The photon bin aggregates contributions from primary photon interactions, as well as secondary photons originating from atomic relaxation and electron-positron annihilation. OpenMC's binning of photon-induced heating was validated by comparing the particle-type heating of a neutron-only and a coupled neutron-photon calculation. When photons were not transported, heating from both electrons and positrons was null, while photon heating matched the combined heating from photons, electrons and positrons of the coupled calculation. This result confirmed that electron and positron heating exclusively stem from photon-induced reactions. Tallies can also be scored according to the material they occur in.

In OpenMC, the study of the coolant heating fraction, expressed as
\begin{equation}
    \zeta = \frac{E_c}{E_{nc} + E_c},
    \label{eqn:FractionEnergyDeposited}
\end{equation}
involves the utilisation of material filters to determine the energy deposited in the coolant, $E_c$, and in non-coolant materials, $E_{nc}$. By applying appropriate particle filters, these quantities can be further decomposed into photonic contribution, denoted by the subscript $\gamma$, and neutronic contribution, represented by the subscript $n$, as shown below:
\begin{equation}
    \begin{gathered}
    E_c = E_{c, \gamma} + E_{c, n}, \\
    E_{nc} = E_{nc, \gamma} + E_{nc, n}.
    \end{gathered}
    \label{eqn:CoolantHeating}
\end{equation}

\section{Results}
\label{sec:Results}
The active region of the model inspired by the \textsl{new}cleo design consists of sintered annular pellets made of a mixed uranium-plutonium oxide (MOX). It is made with slightly depleted natural uranium and plutonium obtained from the reprocessing of spent fuel in PWR reactors in which impurities of Americium are present. At T=20$^\circ$C, each pellet has an outer diameter of 0.714~cm and presents a central cylindrical hollow of 0.2~cm in diameter, coaxial with the pellet itself. The fuel density is 10.403~g/cc at room temperature and the plutonium enrichment is fixed at 25\%. The material used in the cladding tube is an austenitic stainless steel of type 15-15Ti (AIM1) with a density of 7.972~g/cc; the cladding inner diameter is set at 0.737~cm while the outer diameter is 0.85~cm. The pin is filled with helium gas at 2~atm. The distance between the centers of two consecutive fuel rods, i.e. the pitch, is 1.15~cm. The coolant is composed primarily of lead ($^{204}$Pb, $^{206}$Pb, $^{207}$Pb, $^{208}$Pb), with impurities accounting for 0.015\% of the total coolant weight. The fuel and helium temperatures are set at 850$^\circ$C, while the clad and the coolant ones are set at 480$^\circ$C and 470$^\circ$C respectively. The thermal expansion of the fuel and the AIM1 is accounted for using a thermal expansion coefficient determined on the basis of experimental measurements. These yield a density at operating conditions of 10.144~g/cc for the MOX and 7.778~g/cc for the cladding. The density of lead at the operating temperature is set to 10.490~g/cc.

The comparison between the Figures of Merit (FOM) related to the simple pin and lattice geometry for estimating computational performance is presented in Sec.~\ref{sec:FOM}. Focusing specifically on the fraction of energy deposited in lead by various particles, Sec.~\ref{sec:Edep} delves into the influence of bremsstrahlung and examines interactions involving neutrons and photons. The results obtained from a parametric study that varies the coolant temperature and pitch value are presented in Sec.~\ref{sec:param}. Finally, a relation capable of predicting coolant heating in fuel assemblies is proposed in Sec.~\ref{sec:correlation}. Unless indicated otherwise, the simulations comprised 125 batches, each one containing 10$^5$ particles, where the first 25 batches were omitted to ensure the convergence of the source distribution.

\subsection{Calculation setup}\label{sec:FOM}
\begin{figure}[!ht]
    \centering
      \begin{subfigure}{.45\textwidth}
        \centering
        \includegraphics[width=0.7\linewidth]{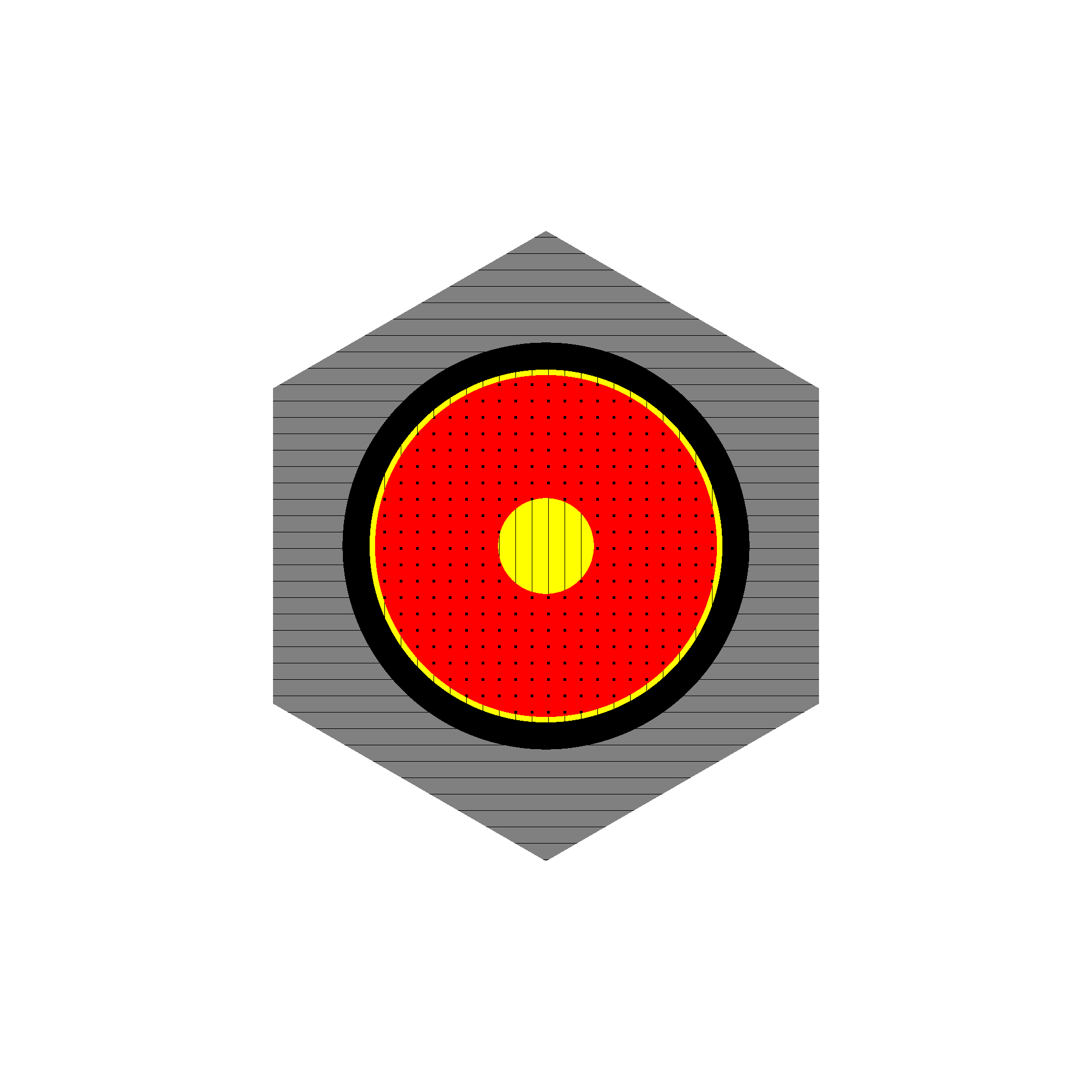}
        \caption{\small Fuel pin cell geometry}
        \label{fig1:sub1}
      \end{subfigure}
      \begin{subfigure}{.45\textwidth}
        \centering
        \includegraphics[width=0.7\linewidth]{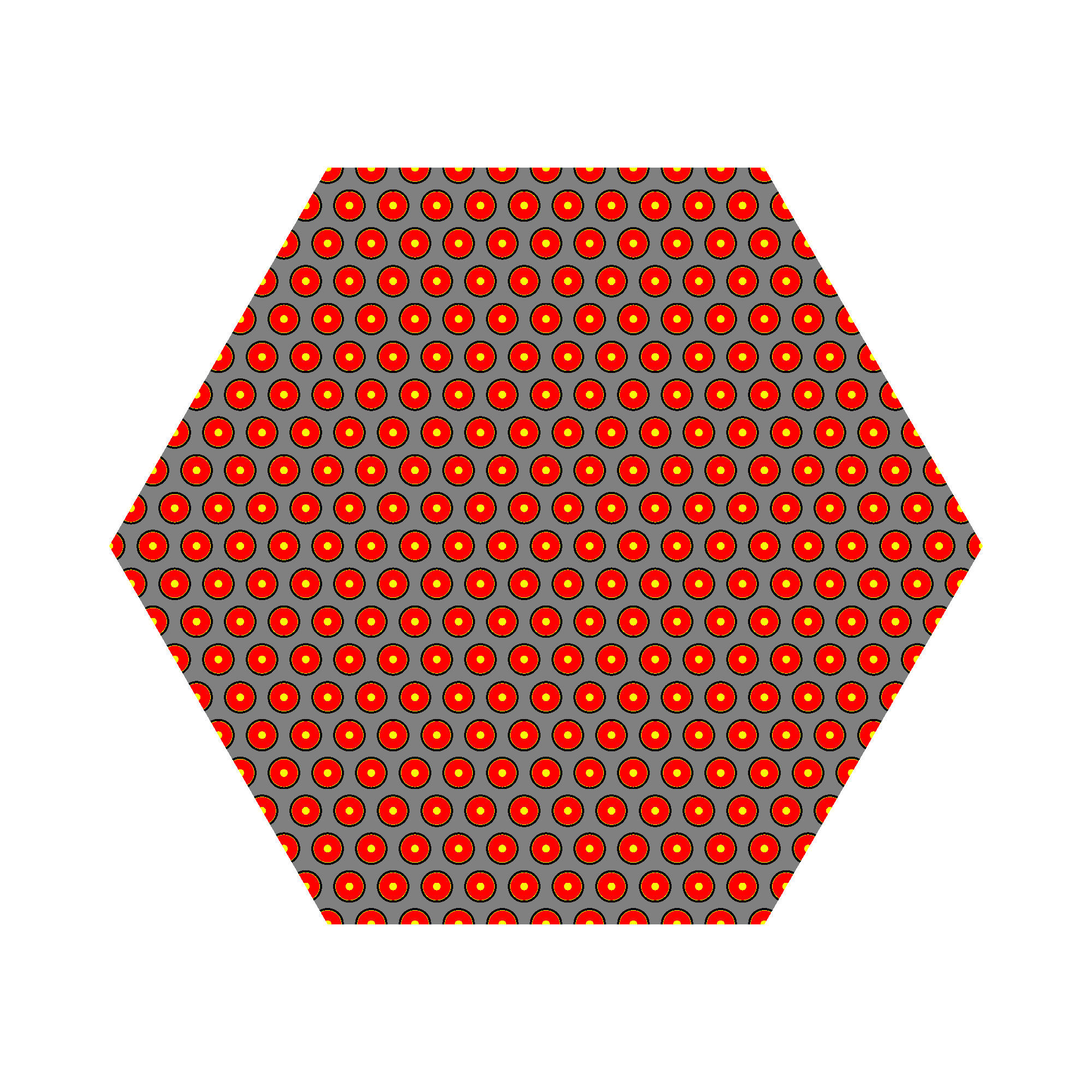}
        \caption{\small Lattice geometry}
        \label{fig1:sub2}
      \end{subfigure}
    \caption{\small OpenMC 2D models for the LFR fuel pin where the fuel is shown in red, the clad in black, the lead coolant in gray and the helium filling in yellow.}
    \label{fig:fig1}
    \end{figure}

A single fuel pin cell configuration would be sufficient to estimate the fraction of energy deposited in the lead coolant. Neutrons in fast reactors travel longer distances between interactions compared to thermal water reactors, which results in a mean free path in LFR of approximately 2-3~cm~\cite{CINOTTI2010}. This quantity was also estimated with OpenMC as the ratio of the tallied flux and total interaction rate resulting in \mbox{$\lambda_{\text{LFR}, n}$ = 3.0846$\pm$0.0047~cm}. For comparison, the same calculation performed for the VERA benchmark Problem 2B~\cite{GODFREY14} yields $\lambda_{\text{PWR}, n}$=1.6050$\pm$0.0012~cm. Given the longer mean free path, the frequent application of reflective boundary conditions in a single fuel pin cell may negatively impact numerical performances. The precision of energy deposition calculations relies on the number of collisions scored in the system volume; if particles bounce between surfaces without interacting, a longer runtime is needed to score enough events to reduce the variance to target values. Implementing a hexagonal lattice, designed with distances between external surfaces greater than $\lambda_{\text{LFR}, n}$, could potentially alleviate this issue. The global performance of the simulations where compared using the figure of merit (FOM) defined as
    \begin{equation}
    	FOM = \frac{1}{\sigma^2 t},
    \end{equation}
where $\sigma$ is the standard deviation of the tally of interest and $t$ is the simulation time~\cite{HAGHIGHAT2020}. Both models can be compared by evaluating the ratio $f = \frac{FOM_{lattice}}{FOM_{pin}}$ where $FOM_{lattice}$ and $FOM_{pin}$ are calculated for the same tally. When $f$ exceeds one, the product between the variance and the simulation time is greater in the pin model compared to the lattice, implying better performance of the latter. The information provided at the beginning of this section was used to implement the 2D geometries representing the fuel pin cell and the lattice. Both are depicted in Fig.\ref{fig:fig1}.

The value for $f$ was determined by varying the lattice size, i.e. changing the number of rings comprising the model, to identify the most efficient system. However, the results for $f$ were not consistent when using the previously chosen settings, necessitating an increase in statistical sampling. Consequently, the analysis was performed considering 100 inactive batches and 400 active ones, each of them containing $10^5$ particles. Results concerning the energy deposition tally are displayed in Tab.~\ref{tab:tab1}.

    \begin{table}[!ht]
    	\centering
    	\caption{\small Calculations of $f$ when varying the number of rings composing the lattice.}
    	\renewcommand{\arraystretch}{1.2}
    	\scalebox{1.05}{
    	\begin{tabular}{cc}\hline
    		rings & $f$ \\ \hline\hline
    		0 & 0.84\\
    		3 & 0.86\\
    		7 & 0.88\\
    		10& 0.89\\
    		\hline
    	\end{tabular}
    	}
    	\label{tab:tab1}
    \end{table}

Since all the values of $f$ from Tab.~\ref{tab:tab1} are less than one, the pin cell geometry proves to be more efficient than the lattice. This behaviour is tied to the treatment of boundary conditions and to the surface-based tracking used by OpenMC~\footnote{\url{https://docs.openmc.org/en/stable/methods/geometry.html} [Accessed: 14/12/2023]}. When reflective conditions are applied, the particle distance to the next collision is not resampled and its direction of flight is altered based on reflection laws. Conversely, the lattice consists of repeated pin cells with transmissive conditions applied, requiring resampling whenever a particle crosses a surface. Because of the long neutron mean free path, this frequent resampling induces a higher computational cost. The performance outcomes could differ when employing alternative tracking techniques in different transport codes.

\subsection{Energy deposition contributions in lead}\label{sec:Edep}
Among the objectives of this study was to assess the energy deposition in lead. This involved analysing energy distribution across the cell's materials and understanding the contributions from the different particle species. To explore the influence of bremsstrahlung, two calculations were performed, each utilising distinct settings for treating electrons (\texttt{led}/\texttt{ttb}). The percentage of total energy deposited in each material and the contribution from particle species are presented in Tab.~\ref{tab:tab2}.
    \begin{table}[!ht]
		\centering
  		\caption{\small Fraction of deposited energy in materials and per particle when considering local energy deposition for electrons and positrons (\texttt{led}) and when allowing for the emission of bremsstrahlung radiation (\texttt{ttb}).}
		\renewcommand{\arraystretch}{1.15}
		\begin{tabular}{cccc}\hline
			& \textbf{Fuel}  [$\%$] & \textbf{Clad} [$\%$] & \textbf{Lead} [$\%$]\\ \hline
			&\multicolumn{3}{c}{\texttt{led}}\\\hline
			Total   & 93.655282 $\pm$ 0.029441& 0.779803 $\pm$ 0.000428&5.564844 $\pm$ 0.002209 \\
			Neutrons & 88.657421 $\pm$ 0.027891 & 0.124885 $\pm$ 0.000053&0.106467 $\pm$ 0.000056  \\
			Photons  &  0.229529 $\pm$ 0.000184&0.016764 $\pm$ 0.000049&0.280495 $\pm$ 0.000209\\
			Electrons & 4.781909 $\pm$ 0.001797&0.636826 $\pm$ 0.000394 &5.187576 $\pm$ 0.002057 \\
			Positrons &  -0.013576 $\pm$ 0.000136& 0.001328 $\pm$ 0.000049  &-0.009694 $\pm$ 0.000162\\
			\hline
			&\multicolumn{3}{c}{\texttt{ttb}}\\\hline
			Total  & 93.643071 $\pm$ 0.035783& 0.783628 $\pm$ 0.000444 &5.573229 $\pm$ 0.002468  \\
			Neutrons &  88.654309 $\pm$ 0.033938& 0.125063 $\pm$ 0.000056 & 0.106648 $\pm$ 0.000059  \\
			Photons   & 0.234145 $\pm$ 0.000172& 0.016817 $\pm$ 0.000049&0.283695 $\pm$ 0.000219\\
			Electrons & 4.780848 $\pm$ 0.001938& 0.640930 $\pm$ 0.000401 & 5.207398 $\pm$ 0.002295\\
			Positrons & -0.026231 $\pm$ 0.000121 & 0.000818 $\pm$ 0.000041 & -0.024511 $\pm$ 0.000148\\
			\hline
		\end{tabular}
        \label{tab:tab2}
	\end{table}

As expected, the calculations highlighted that the majority of energy deposition occurs in the fuel. This is consistent with the fact that most of the energy released from fission is associated with the kinetic energy of fission fragments, which is deposited locally. The importance of neutron-induced reactions in the fuel is underlined by the predominant fraction of energy deposition coming from neutrons (88.6\%). Regarding the energy deposited in the coolant, the lead heating fraction is approximately 5.6\%, largely due to photon-induced reactions. As explained in Sec.~\ref{ssec:openmc}, these reactions include the contribution from both electron and positron heating.

The total energy deposition fraction was consistent between the \texttt{led} and \texttt{ttb} calculations (Welsh test: t$\leq$1).
Switching from the \texttt{led} to \texttt{ttb} option revealed a small increase in the energy deposition of photons and electrons in lead, while a decrease in the fraction of energy deposition related to positrons was observed. This effect correlates with an augmented photon count in the \texttt{ttb} calculation: bremsstrahlung photons contribute to photon heating along with other secondary photon processes. Positrons partake in this process, therefore their energy deposition fraction turns increasingly negative from \texttt{led} to \texttt{ttb}. Their energy loss benefits photons and electrons: secondary photons possess lower energy than the initially generated ones and interact predominantly through the photoelectric effect, leading to increased electron emission and initiating a cycle of photon-electron production, as electrons also emit bremsstrahlung photons. When the \texttt{led} electron treatment is selected, all energy linked to electrons and positrons is deposited in matter, preventing the increase of photons through bremsstrahlung and impacting their energy deposition fraction. This behaviour is corroborated by the photon interaction rates depicted in Tab.~\ref{tab:tab3}: the \texttt{ttb} setting showcases an increase in photon count and, consequently, an increase in the number of reactions they participate in.

    \begin{table}[!ht]
	\centering
	\caption{\small Photons reaction rates in lead with the \texttt{led} and \texttt{ttb} options.}
	\renewcommand{\arraystretch}{1.2}
    	\begin{tabular}{ccc}\hline
    		\multicolumn{3}{c}{Photons interaction rate in lead [$10^{17}/s$]}\\\hline\hline
    		& \texttt{led} &  \texttt{ttb} \\ \hline
    		Rayleigh & 2.0326 $\pm$ 0.0005  & 2.6706 $\pm$ 0.0008  \\
    		Compton &  8.7168 $\pm$ 0.0025 &  9.1685 $\pm$ 0.0029\\
    		Photoelectric & 22.8279 $\pm$ 0.0058 & 38.4725 $\pm$ 0.0120 \\
    		Pair production & 0.4786 $\pm$ 0.0002 &  0.4807 $\pm$ 0.0002  \\\hline
    		
    	\end{tabular}
	\label{tab:tab3}
    \end{table}

The significance of the photoelectric effect elucidates why, when considered individually, electrons constitute the primary contributor to lead heating. However, the marginal disparity in magnitude ($\leq0.1$\%) between the two sets of results signifies that bremsstrahlung plays a minor role compared to Coulomb's interaction in the energy deposition of charged particles.

\begin{figure}[!ht]
        \centering
        \includegraphics[width=0.9\linewidth]{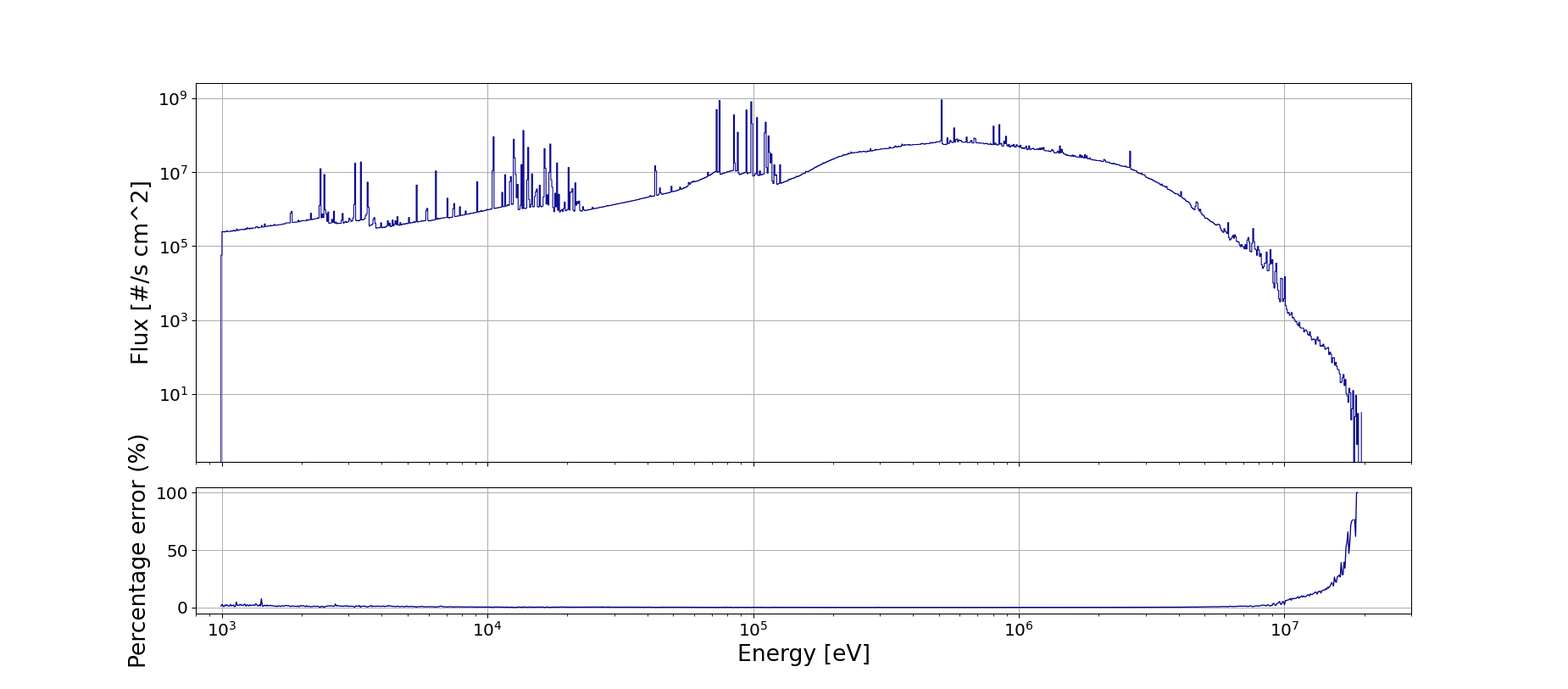}
        \caption{\small Photon spectrum along with its percentage uncertainty computed using OpenMC.}
        \label{fig:photon_spectrum}
    \end{figure}

The importance of the photoelectric effect is reaffirmed by the photon spectrum analysed within the geometry volume. Figure~\ref{fig:photon_spectrum} depicts the photon flux derived from a \texttt{ttb} calculation \footnote{The calculation involved 500 batches, 100 inactive, each comprising $5\cdot10^{5}$ particles, using the ECCO 2000-group structure~\cite{sartori1990standard}.}, plotted against particle energy, together with the percentage error associated with the calculations. It is noticeable that the majority of particles occupy the energy range below 1~MeV, dominated by the photoelectric effect. Distinctive sharp discontinuities characteristic of this process are evident, alongside peaks linked to scattering events, which represent the second most significant photon interaction.

\begin{figure}[!ht]
        \centering
        \includegraphics[width=0.9\linewidth]{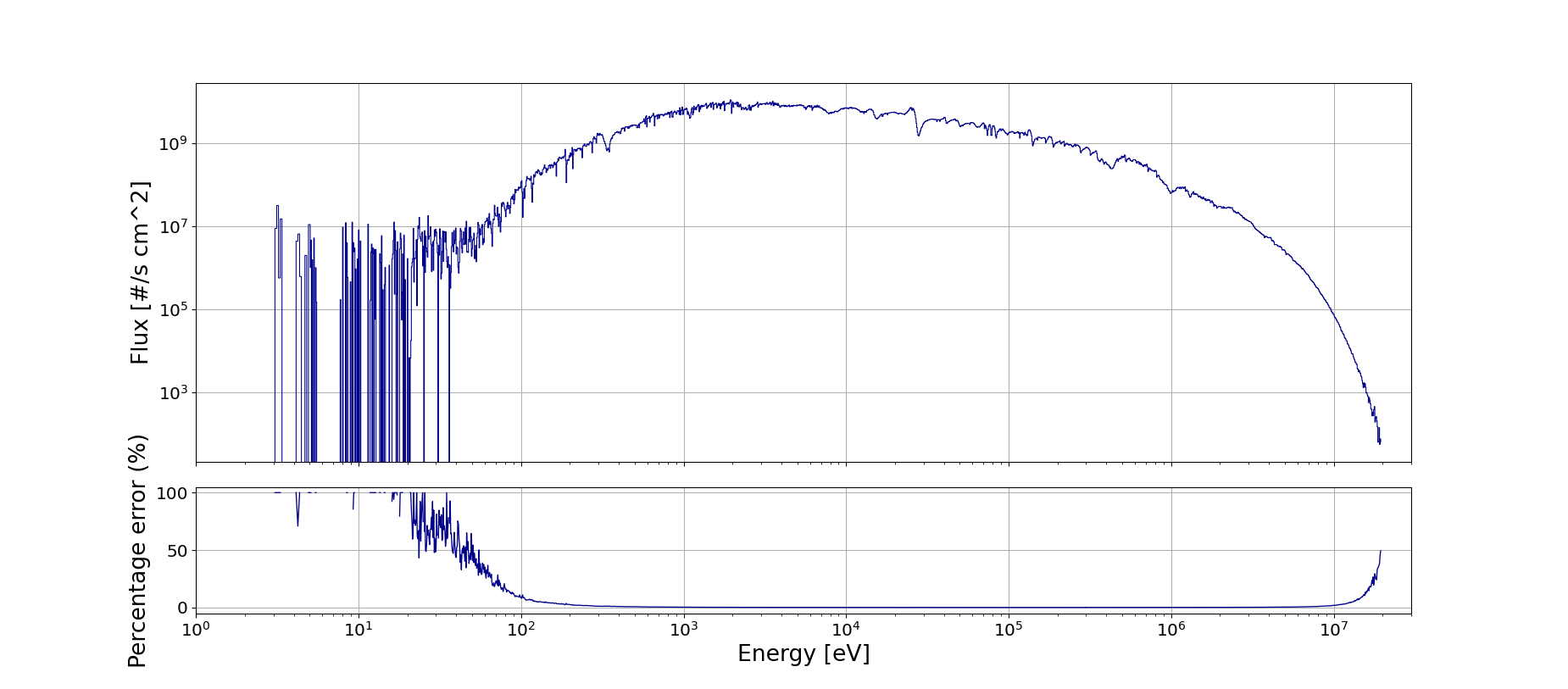}
        \caption{\small Neutron spectrum along with its percentage uncertainty computed using OpenMC.}
        \label{fig:neutron_spectrum}
    \end{figure}

Identifying neutron-related processes in lead proved more straightforward. Elastic scattering was the predominant interaction while inelastic scattering was less than 1\%. This observation aligns with the neutron spectrum within the LFR volume, represented in Fig.~\ref{fig:neutron_spectrum}, as the most populated region corresponds to the energy range where elastic scattering is dominant. Despite the large number of collisions in lead, the minimal energy loss experienced by neutrons per elastic scattering event translated to their modest contribution of approximately $2$\% to the total coolant heating. Absorption processes in the coolant were also minimal, constituting around $0.05$\% of total events, with radiative capture being the prevalent channel (99.98\%).

\subsection{Parametric study}\label{sec:param}
Considering the ongoing research on LFR fuel element design and the anticipation of diverse operational conditions, it becomes essential to examine how variations in pitch value and coolant temperature might impact heating in lead.

The lead temperature directly affects its density, subsequently influencing the macroscopic cross section values used in calculations. To assess potential variations, the analysis of energy deposition with the \texttt{ttb} option was repeated for two supplementary coolant temperatures: 430$^\circ$C and 510$^\circ$C, representing the inlet and outlet temperatures respectively. These values are representative of different power operation regimes where molten lead is always in subcooled conditions and far from saturation.
Results concerning the energy deposition distribution in materials are displayed in Tab.~\ref{tab:tab5}. 

    \begin{table}[!ht]
    	\centering
    	\caption{\small Total energy deposition percentage in materials at three different coolant temperatures.}
    	\renewcommand{\arraystretch}{1.25}
    	\scalebox{1.05}{
    		\begin{tabular}{cccc}\hline
    			& 430°C &  470°C & 510°C \\ \hline\hline
    			\multicolumn{4}{c}{Total energy deposition [$\%$]}\\\hline
    			Fuel  & 93.631734 $\pm$ 0.029862 & 93.643071 $\pm$ 0.035783 &93.661990 $\pm$ 0.031503\\
    			Helium& 0.000069 $\pm$ 0.000002  & 0.000073 $\pm$ 0.000002 &0.000074 $\pm$ 0.000002 \\
    			Clad  & 0.782029 $\pm$ 0.000389  & 0.783628 $\pm$ 0.000444 &0.784472 $\pm$ 0.000434\\
    			Lead  & 5.586168 $\pm$ 0.002126  & 5.573229 $\pm$ 0.002468 &5.553464 $\pm$ 0.002278\\\hline
    		\end{tabular}
    	}
    	\label{tab:tab5}
    \end{table}

The differences in heating fractions were less than 1\%, mirroring similar trends in the values of reaction rates. These findings suggest that variations in coolant temperature have an insignificant impact on the energy deposited in lead. This is supported by the small variation of lead density with temperature, as demonstrated in Tab.~\ref{tab:lead-density}, which implies that changes to the macroscopic cross sections due to changes in the coolant temperature can be neglected in this study. Different outcome was found for pressurized water reactors in~\cite{CASTAGNA}.

    \begin{table}[!ht]
    	\centering
    \caption{\small Lead density when heated at three different temperatures.}
    	\renewcommand{\arraystretch}{1.2}
    	\scalebox{1.02}{
    	\begin{tabular}{cccc}\hline
    		&430°C &  470°C& 510°C \\ \hline
    		density [$g/cc$]& 10.5413  &10.4901 &10.4389  \\\hline
    	\end{tabular}
    	}
    \label{tab:lead-density}
    \end{table}

\begin{figure}[!ht]
    \centering
      \begin{subfigure}{.45\textwidth}
        \centering
        \includegraphics[width=1\linewidth]{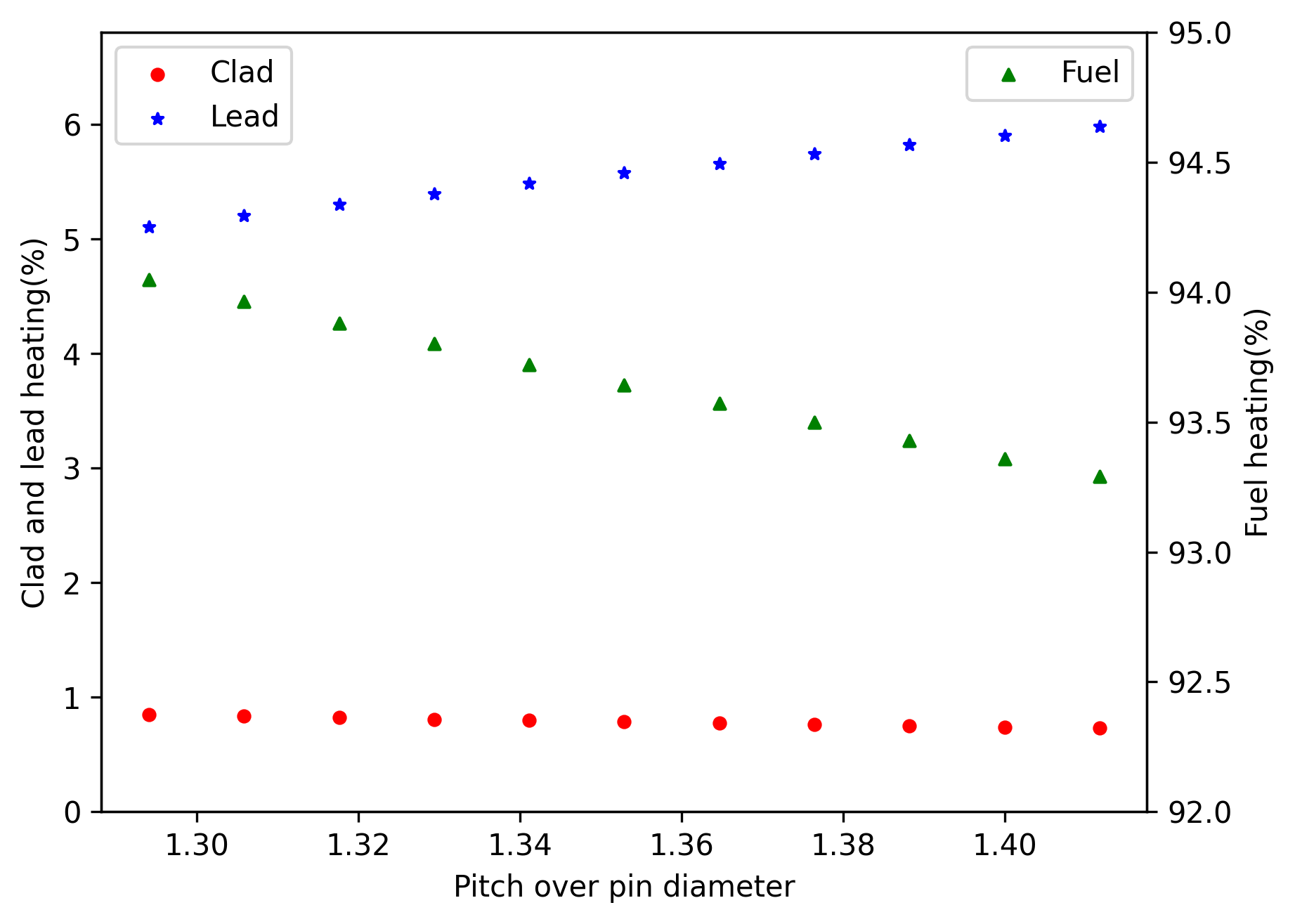}
        \caption{}
        \label{fig3:sub1}
      \end{subfigure}
      \begin{subfigure}{.45\textwidth}
        \centering
        \includegraphics[width=1\linewidth]{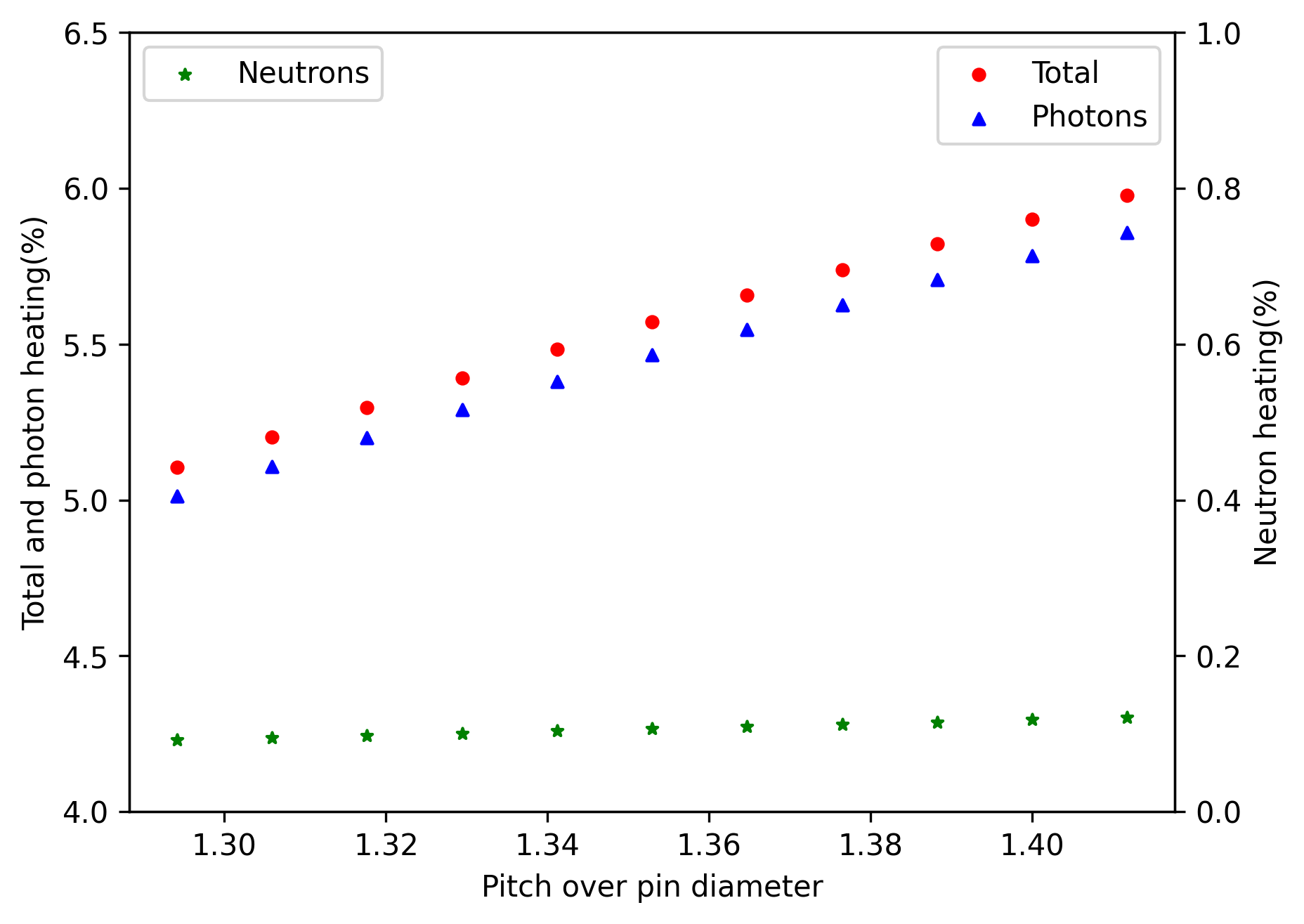}
        \caption{}
        \label{fig3:sub2}
      \end{subfigure}
    \caption{\small Total energy deposition fractions in materials (a) and in lead (b) as a function of pitch over pin diameter where the latter was kept fixed at 0.85~cm.}
    \label{fig:fig3}
    \end{figure}

To investigate the influence of the pitch value on energy deposition, the previous calculation was replicated considering pitch values ranging from $1.10$~cm to $1.20$~cm in increments of $0.01$~cm. The resulting total energy deposition for each material is plotted in Fig.~\ref{fig3:sub1}.
From the plot, the fraction of energy deposited in the cladding remains relatively constant, whereas it varies in fuel and lead. As energy deposition in fuel decreases, the fraction in lead increases from 5.10$\%$ to 5.98$\%$. This behaviour appears to be associated with a mass-driven effect, wherein enlarging the pitch expands the quantity of lead that particles can interact with and be absorbed by. This aspect was further investigated by analysing particle species' contribution to energy deposition in lead, as illustrated in Fig.~\ref{fig3:sub2}. Photons were confirmed as the primary contributor to energy deposition due to their short mean free path $\lambda_{\text{LFR}, \gamma} = 0.3374\pm0.0001$~cm, resulting in the complete loss of their energy in the coolant. The increase in lead mass is mirrored by an increase in reaction rate in lead for both photons and neutrons as depicted in Tab.~\ref{tab:tab6}, reinforcing the mass effect hypothesis. However, it is the increase in photon interaction rates which explains the greater fraction of energy deposited.

\begin{table}[!ht]
	\centering
	\caption{\small Neutron and photon reaction rates in lead for different pitch over pin diameter values, as expressed in percentage difference from the preceding pitch value.}
	\renewcommand{\arraystretch}{1.13}
    \makebox[1 \textwidth][c]{       
    \resizebox{1\textwidth}{!}{
        \begin{tabular}{cccccccccccc}\hline
            & $1.29$ & $1.31$ & $1.32$ & $1.33$ & $1.34$ & $1.35$ & $1.36$ & $1.38$ & $1.39$ & $1.40$
            & $1.41$ \\\hline
            \multicolumn{12}{c}{Neutron interactions}\\\hline
            Scattering & -- & 4.02(8) & 3.90(8) & 3.78(8) & 3.68(8) & 3.59(8) & 3.48(8) &3.41(7) & 3.31(8) & 3.24(7) & 3.17(7)\\
            Absorption & -- & 3.86(18) & 3.83(18) & 3.75(16) & 3.49(16) & 3.60(15) & 3.40(14) & 3.33(16) & 3.28(16) & 3.11(16) & 3.14(15)\\ 
            \hline    
            \multicolumn{12}{c}{Photons interactions}\\\hline
            Rayleigh & -- & 1.75(8)  &1.67(8)  &1.62(7)  & 1.53(7) &1.48(8) &1.36(8)&1.34(7)&1.31(7)& 1.25(8)&1.21(8) \\
            Compton & -- & 1.89(8) & 1.81(8) & 1.74(7) & 1.69(7) & 1.60(8)& 1.48(8)& 1.43(8)& 1.44(8)& 1.35(8) & 1.29(8)\\
            Photoelectric& -- & 1.71(8) & 1.63(8) & 1.58(8) & 1.51(7) & 1.44(8)& 1.33(8)& 1.30(7)& 1.30(7)& 1.23(8)& 1.18(8)\\
            Pair production & -- & 1.93(13) & 1.83(13) & 1.79(11) & 1.74(11) & 1.67(12)& 1.51(13) & 1.44(12) & 1.49(11)& 1.36(12)& 1.33(13)\\
            \hline
        \end{tabular}
    }
    }
	\label{tab:tab6}
\end{table}

\subsection{Energy deposition correlation function in LFR fuel assemblies}\label{sec:correlation}
Studying the energy deposition in lead revealed the predominant influence of the geometrical configuration on the fraction of coolant heating. The goal was to formulate a relation capable of predicting the fraction of energy deposited in the coolant in a fuel assembly (FA) through simple fuel pin calculations and based on geometrical parameters. The correlation function was devised by considering two different fuel assemblies: one from the ALFRED benchmark~\cite{ALEMBERTI2014} and the other from the ongoing developments of the LFR-30 reactor at \textsl{new}cleo. In the latter, the fuel bundle, made out of the same fuel described at the beginning of Sec.~\ref{sec:Results}, is enclosed by an inner and outer hexagonal wrapper of thickness 0.25~cm, made from AIM1 steel. The outer keys of these wrappers measure 19.7~cm and 6.80~cm respectively. The bundle consists of 234 fuel pins arranged in a hexagonal lattice of pitch 1.15~cm. The space enclosed by the inner wrapper is filled with lead. Additionally, an outer layer of lead of thickness 0.30~cm surrounds the outer wrapper.

\begin{figure}[htbp]
  \centering
  \begin{subfigure}{0.23\textwidth}
    \centering
    \includegraphics[width=\textwidth]{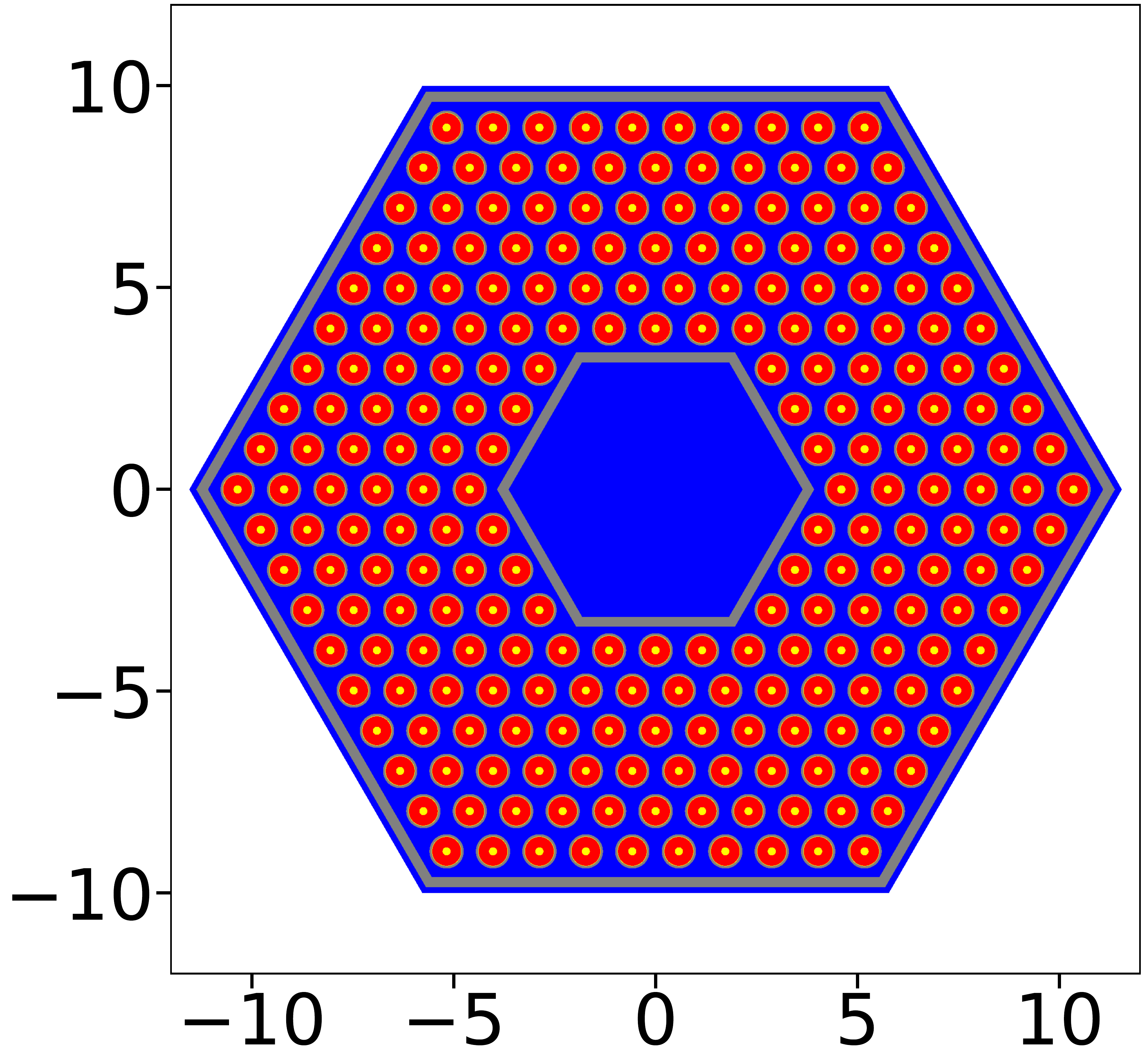}
    \caption{}
    \label{fig:LFROriginal}
  \end{subfigure}
  \begin{subfigure}{0.23\textwidth}
    \centering
    \includegraphics[width=\textwidth]{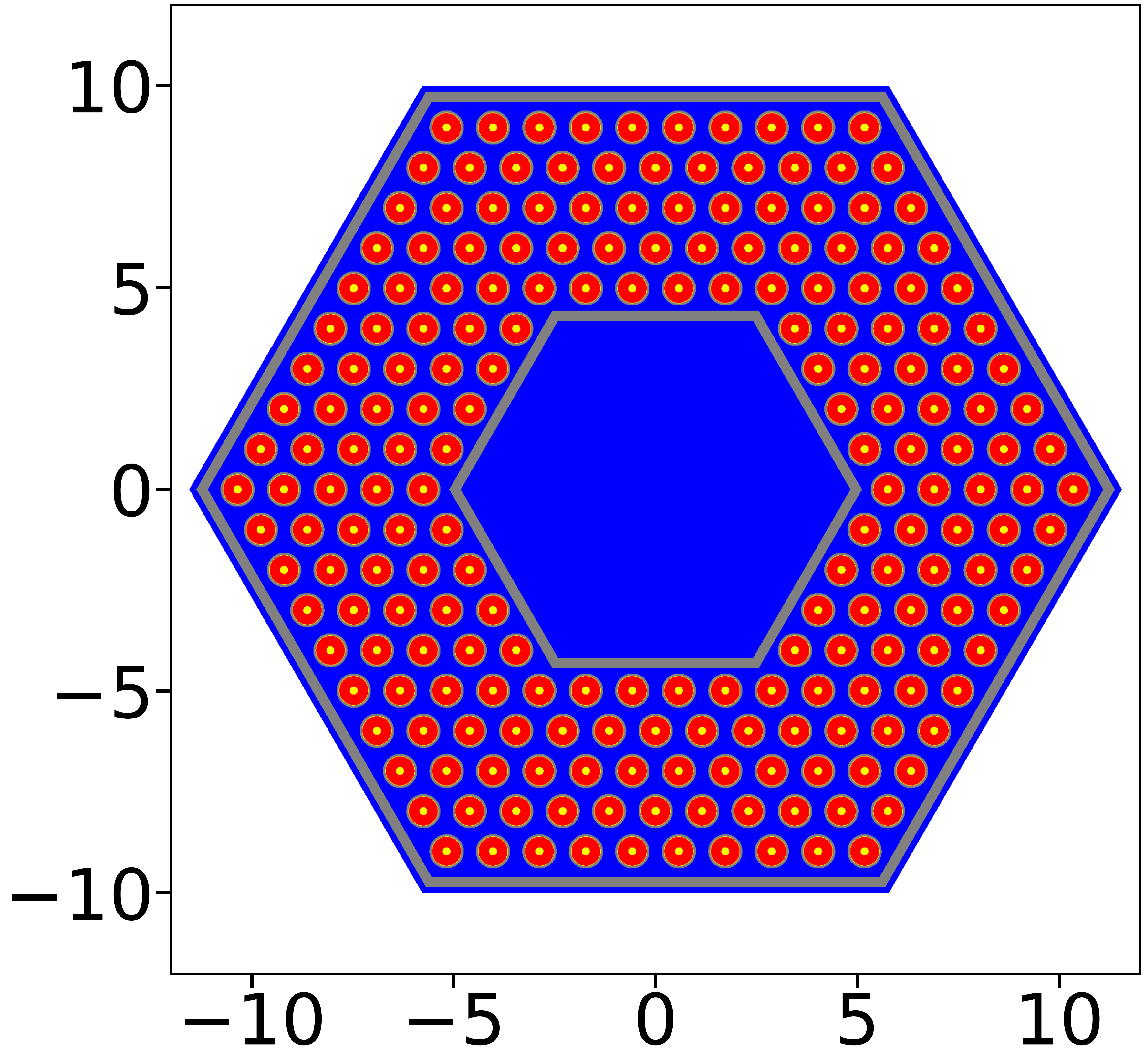}
    \caption{}
    \label{fig:LFRConf1}
  \end{subfigure}
  \begin{subfigure}{0.23\textwidth}
    \centering
    \includegraphics[width=\textwidth]{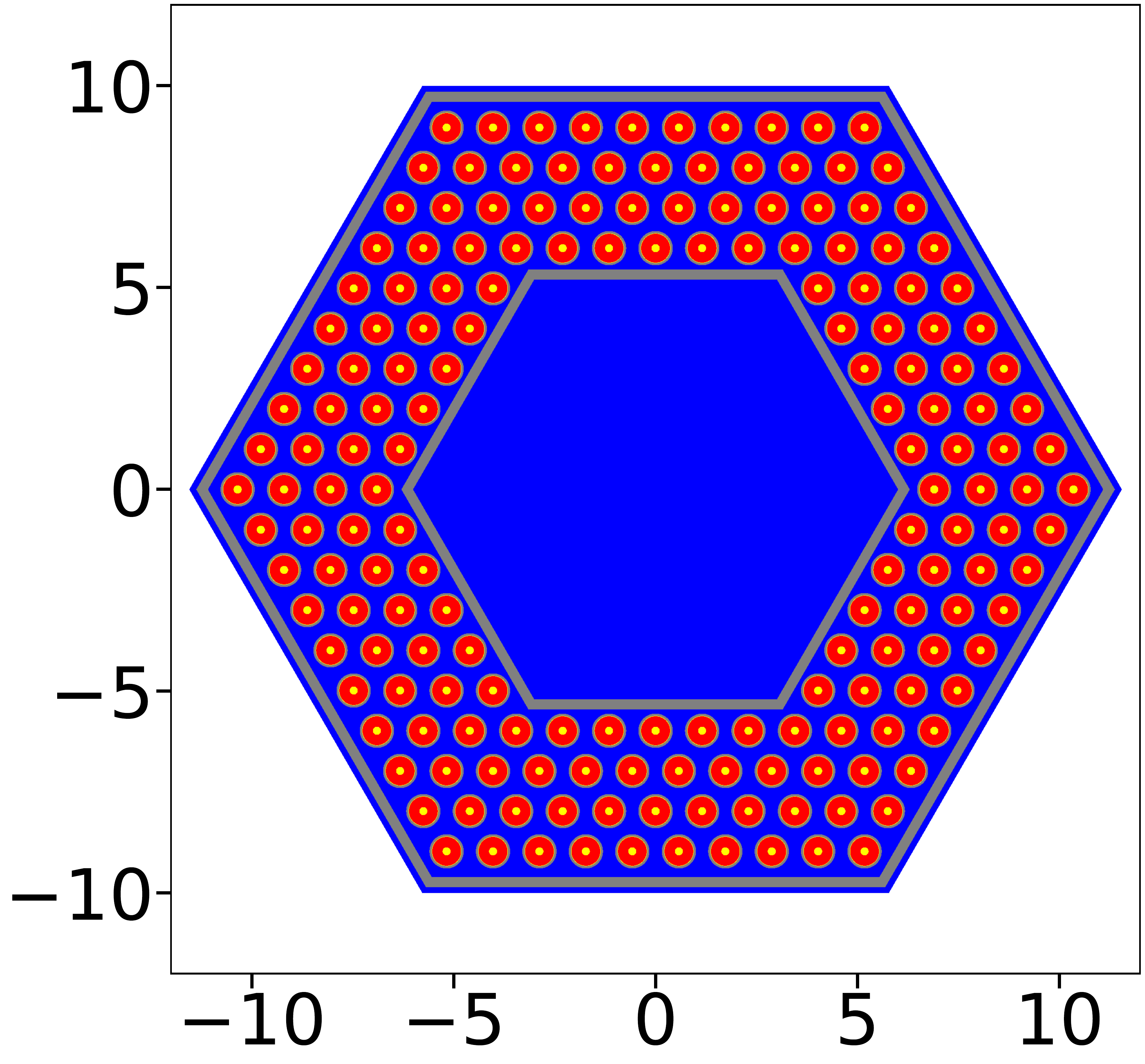}
    \caption{}
    \label{fig:LFRConf2}
  \end{subfigure}
  \begin{subfigure}{0.23\textwidth}
    \centering
    \includegraphics[width=\textwidth]{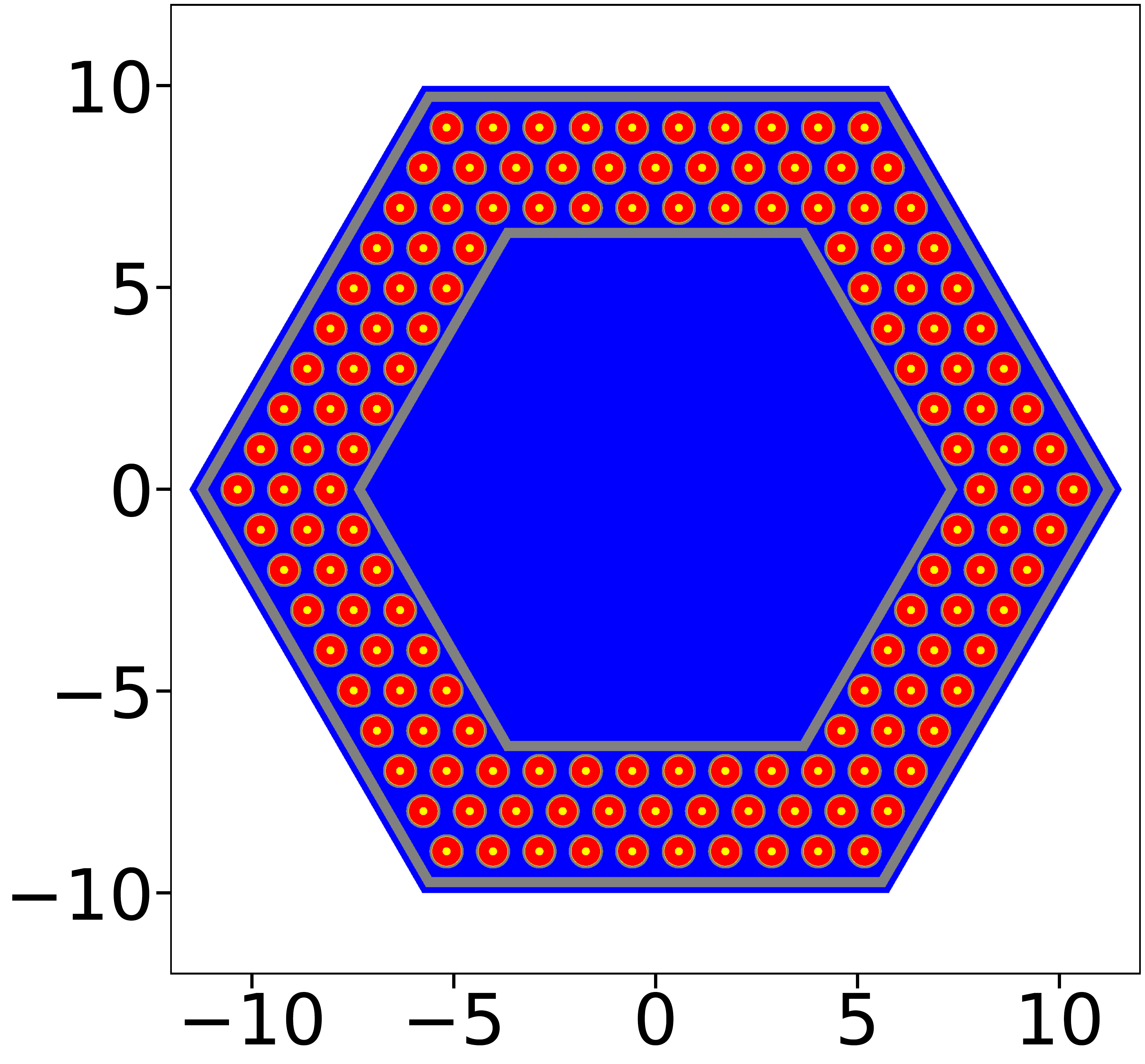}
    \caption{}
    \label{fig:LFRConf3}
  \end{subfigure}
  \caption{\small Original (a) and additional three modified configurations (b, c, d) of the LFR-30 fuel assembly, visualised using OpenMC. The outer key of the inner channel is 6.80~cm, 8.85~cm, 10.90~cm and 12.95~cm respectively. MOX fuel is highlighted in red, AIM1 steel in grey, lead in blue and helium in yellow.}
  \label{fig:LFRConfigs}
\end{figure}

\begin{figure}[htbp]
  \centering
  \begin{subfigure}{0.23\textwidth}
    \includegraphics[width=\textwidth]{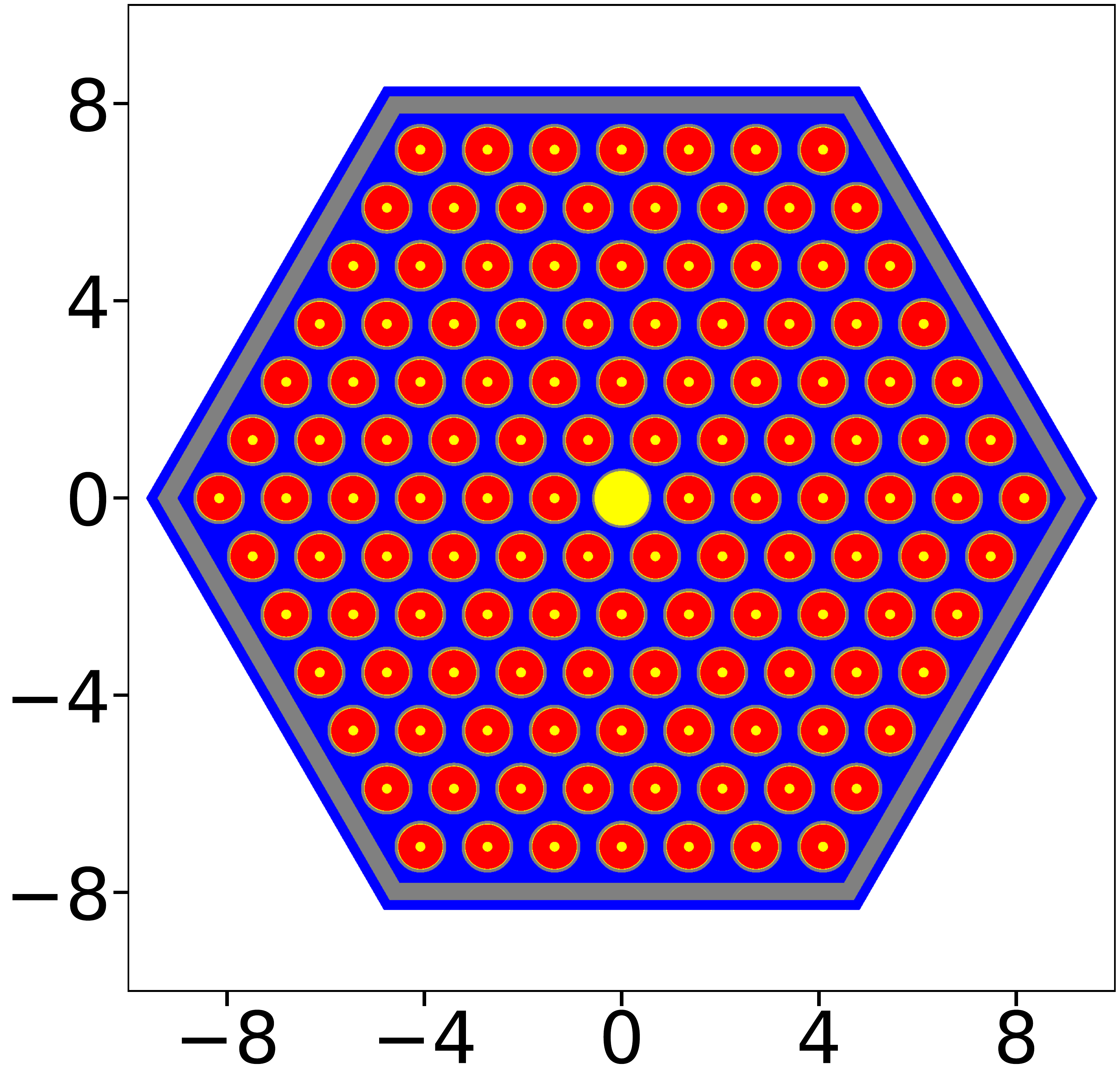}
    \caption{}
    \label{fig:ALFREDOriginal}
  \end{subfigure}
  \begin{subfigure}{0.23\textwidth}
    \includegraphics[width=\textwidth]{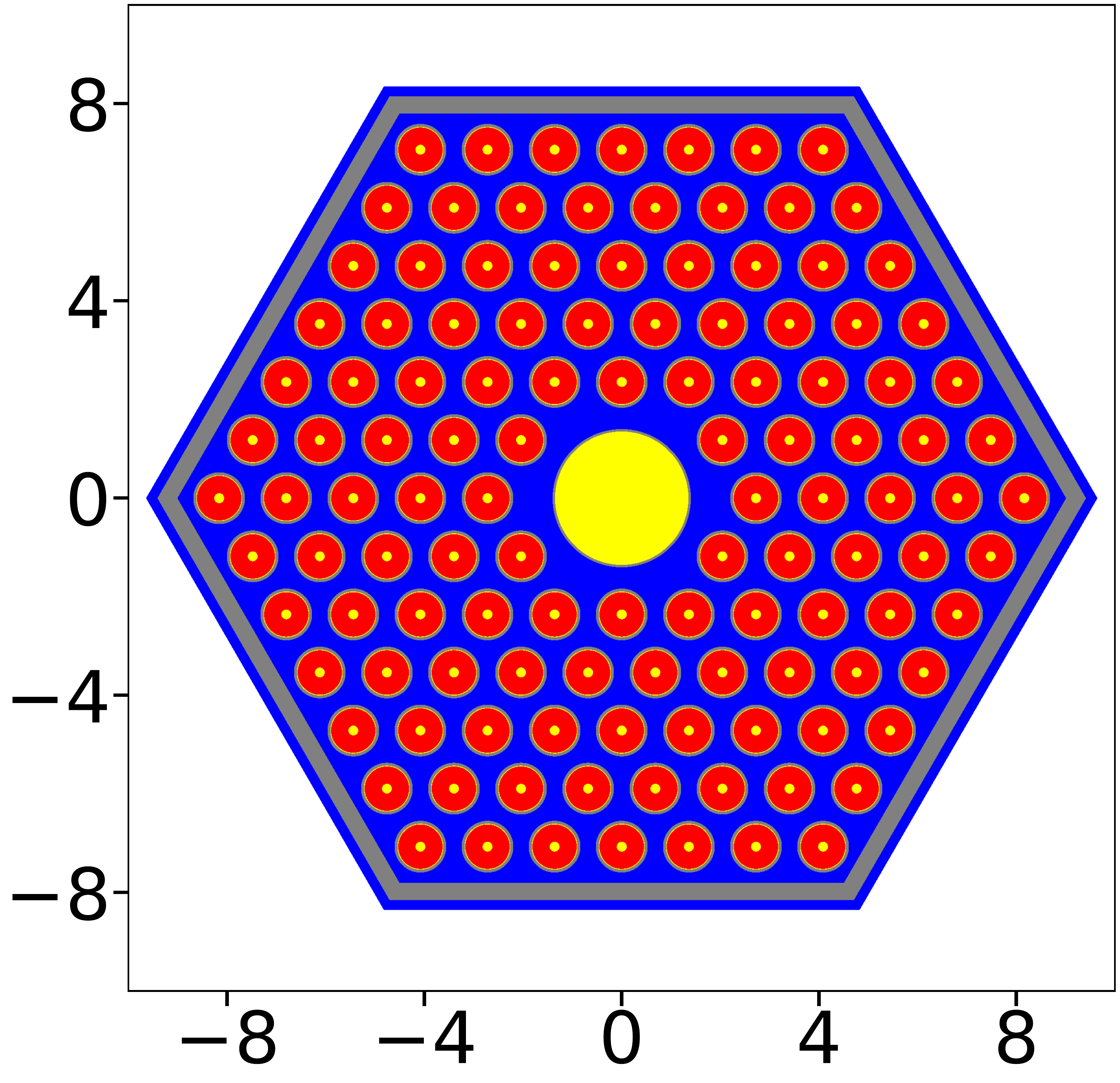}
    \caption{}
    \label{fig:ALFREDConf1}
  \end{subfigure}
  \begin{subfigure}{0.23\textwidth}
    \includegraphics[width=\textwidth]{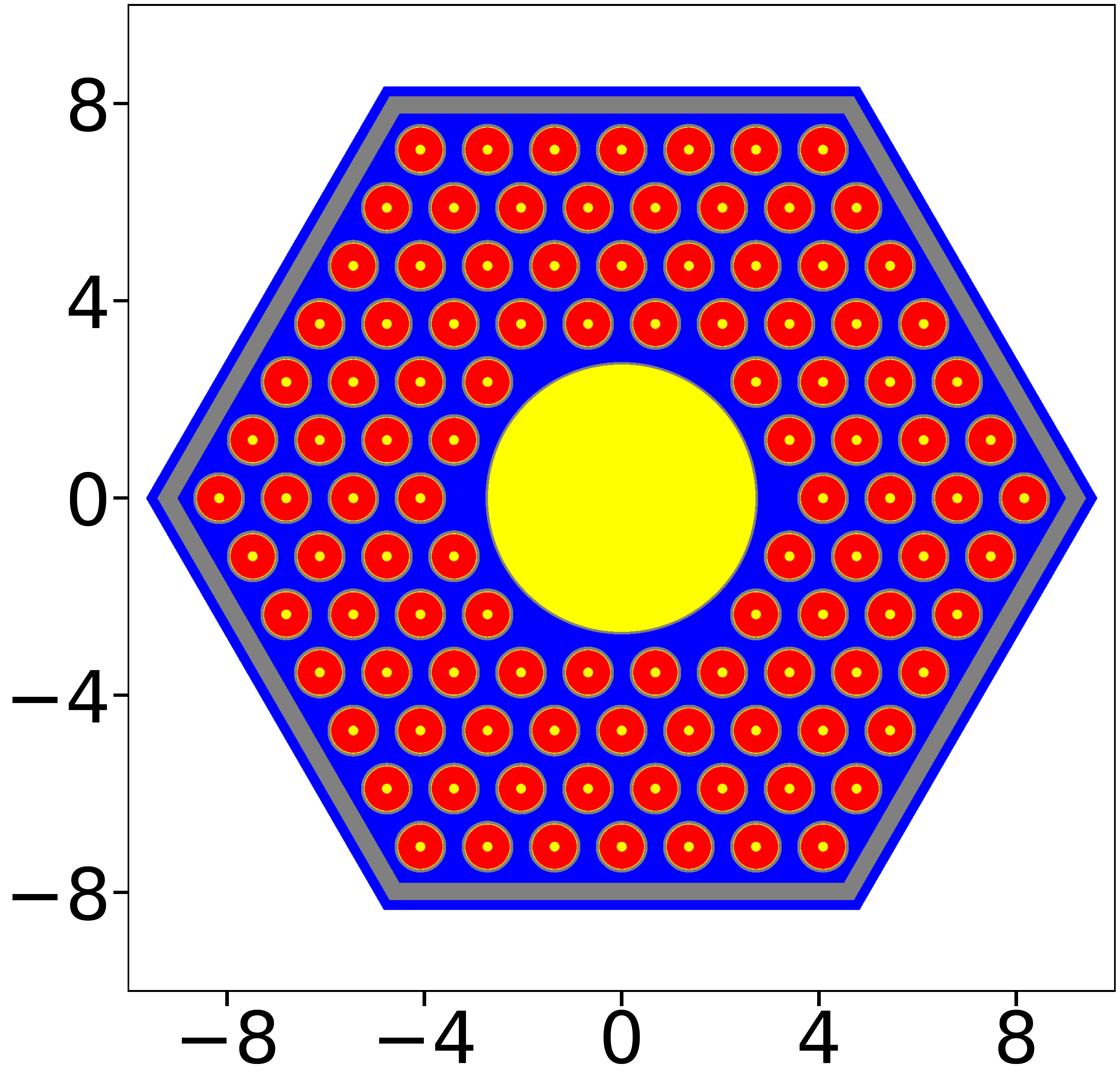}
    \caption{}
    \label{fig:ALFREDConf2}
  \end{subfigure}
  \begin{subfigure}{0.23\textwidth}
    \includegraphics[width=\textwidth]{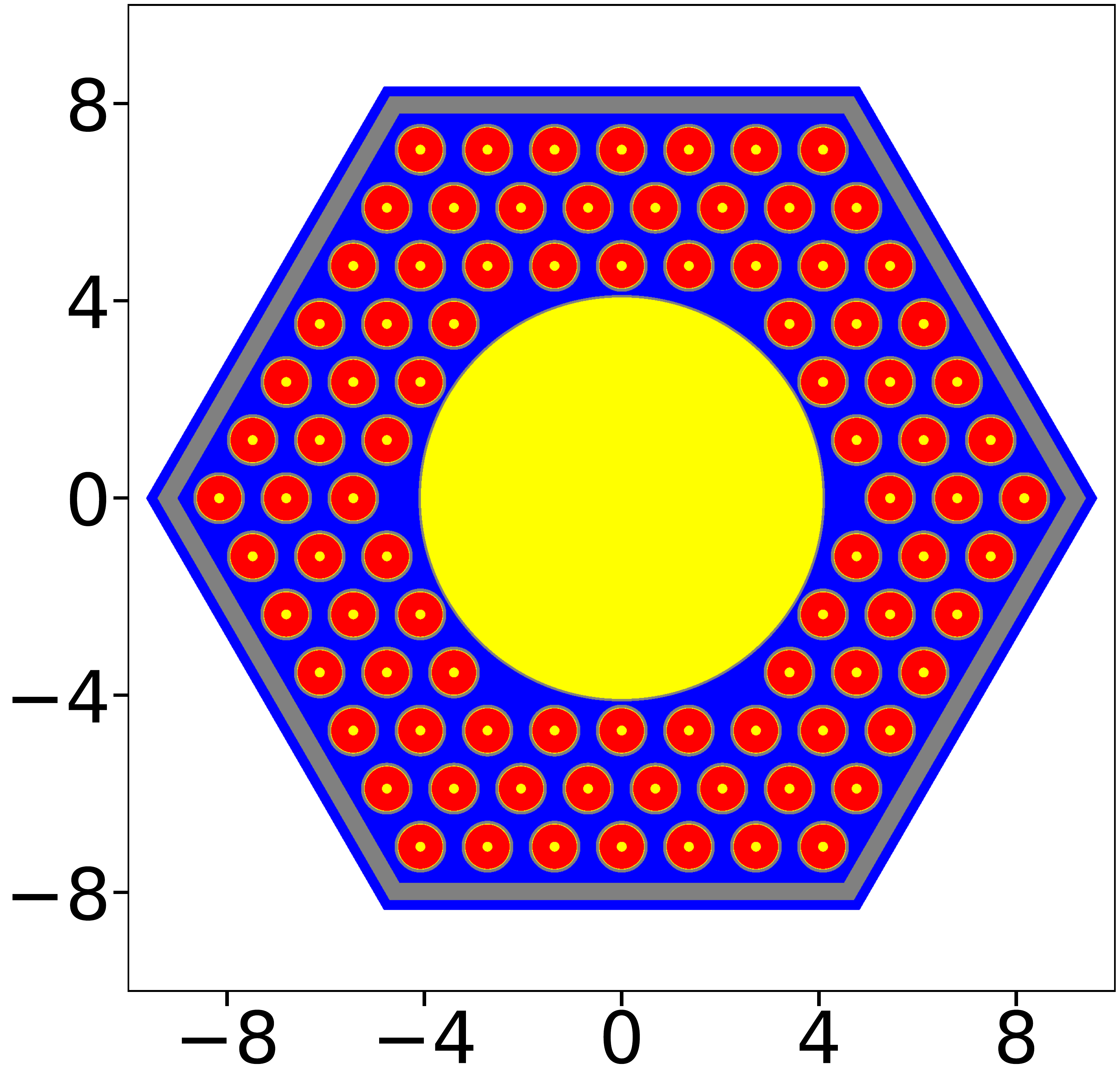}
    \caption{\small}
    \label{fig:ALFREDConf3}
  \end{subfigure}

  \caption{\small Original (a) and additional three modified configurations (b, c, d) of the ALFRED fuel assembly, visualised using OpenMC. The radius of the central dummy pin is 0.6~cm, 1.4~cm, 2.76~cm and 4.12~cm respectively. MOX fuel is highlighted in red, AIM1 steel in grey, lead in blue and helium in yellow.}
  \label{fig:ALFREDConfigs}
\end{figure}

Three additional assembly configurations were derived from the original FAs to assess the relation's validity in different emphasised designs. These were created by enlarging the inner hexagonal channel filled with lead for the LFR-30 fuel assembly, illustrated in Fig.~\ref{fig:LFRConfigs}, and by expanding the size of the circular channel filled with helium for the ALFRED model, presented in Fig.~\ref{fig:ALFREDConfigs}. Although extreme and abstract, these configurations serve the purpose of studying the physics of coolant heating in the LFR fuel elements.

As demonstrated in Sec.~\ref{sec:Edep}, more than 97\% of the energy deposited in lead is photon-related heating. Neutrons, have a small mass compared to lead atoms and interact 99\% of the time through elastic scattering, depositing only small fractions of energy at each collision event. Thus, the coolant heating in lead from Eqn.~\ref{eqn:CoolantHeating} can be approximated as
\begin{equation}
    E_c \simeq E_{c, \gamma}.
\end{equation}

\begin{figure}[htbp]
    \centering
    \includegraphics[width=\textwidth]{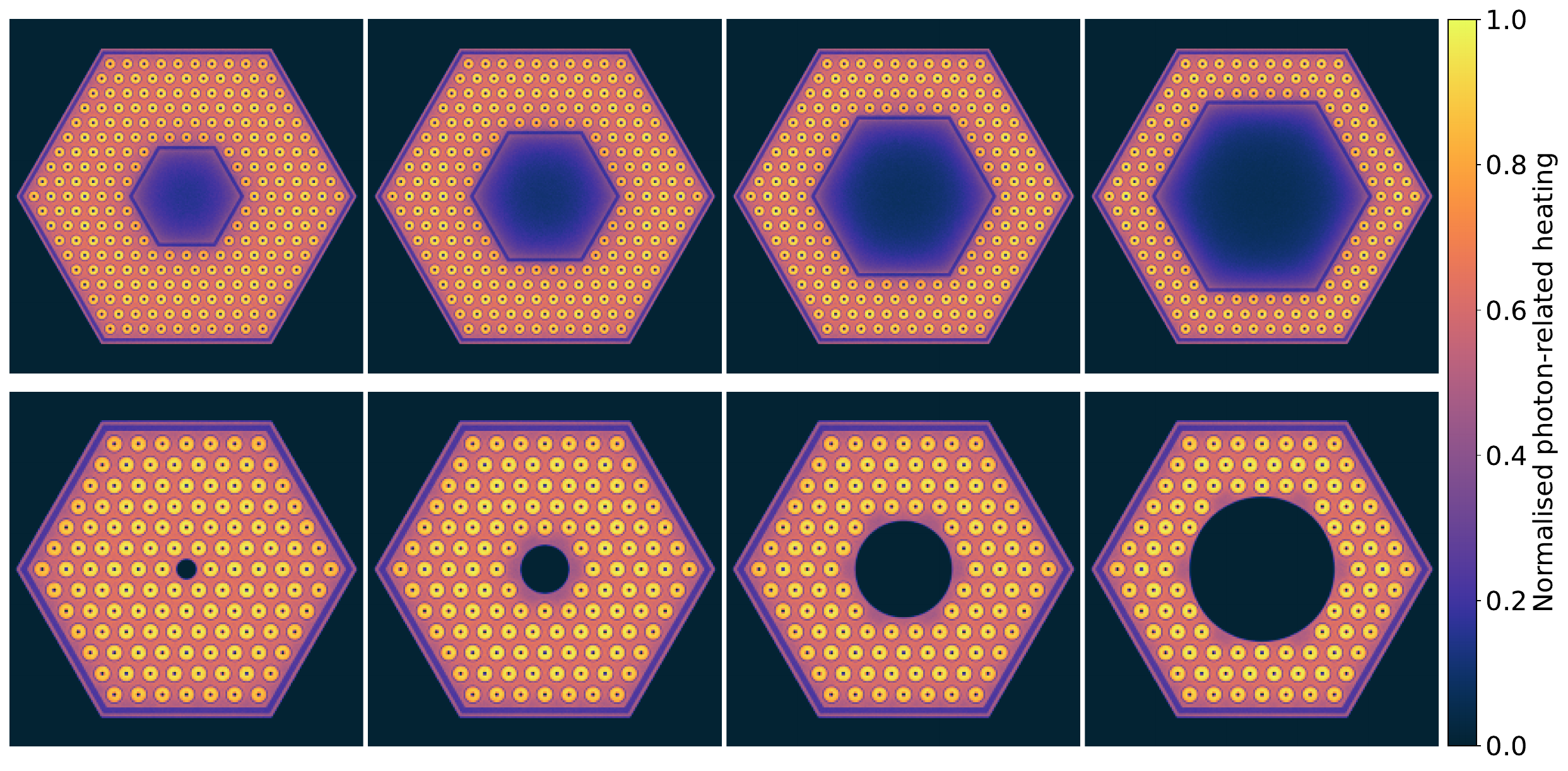}
    \caption{\small Normalised photon-related heating across the LFR-30 and ALFRED fuel assembly configurations in the top and bottom row respectively.}
    \label{fig:2DHeatingLFR30}
\end{figure}

The next step was to identify a relation between the photon-related coolant heating in the assembly, $E_{c, \gamma, A}$, and that in the pin cell, $E_{c, \gamma, FC}$. As suggested in Sec.~\ref{sec:param}, the initial hypothesis was to consider that the energy deposited in the coolant is proportional to the volume of lead in the assembly, assuming that the assembly has the same heating density as the pin cell. However, the distribution of photon-related heating across the assemblies, plotted in Fig.~\ref{fig:2DHeatingLFR30}, revealed that volumes of lead on the periphery, and in the central channel for the LFR assembly, experienced significantly lower levels of heating compared to lead inside the lattice of fuel pins. The high opacity of lead due to the small mean free path of photons in that medium, $\lambda_{\text{LFR}, \gamma} = 0.3348\pm0.0001$ \unit{cm} across all assembly configurations, explains this behaviour.

To account for this phenomenon, an approximation to $E_{c, \gamma, A}$ was formulated by separating the total volume of lead into two parts: the volume surrounding the fuel pins and the remaining volume, and by scaling them with different heating densities. The volume surrounding the fuel pins, equal to $N_{FC}V_{c, FC}$ - where $N_{FC}$ is the number of fuel pins in the assembly and $V_{c, FC}$ is the volume of lead in the fuel cell - was multiplied by the heating density obtained from the pin cell calculations while the extra volume - computed as $V_{c, \text{extra}, A} = V_{c,A} - N_{FC}V_{c,FC}$ - was scaled by the pin cell's heating density reduced by a penalty factor $g$. In essence, the product $gV_{c, \text{extra}, A}$ can also be viewed as the additional volume of lead in the assembly that contributes to the heating. The approximation can be expressed as follows:
\begin{equation}
    \begin{split}
    E_{c, \gamma, A}    & = E_{c,\gamma,FC} N_{FC} \left(1 + g\frac{V_{c, \text{extra}, A}}{N_{FC}V_{c,  FC}}\right) \\
                        & = E_{c,\gamma,FC} N_{FC} \kappa,
    \end{split}
    \label{eqn:PhotonRelatedHeating}
\end{equation}
where the term in brackets is defined as $\kappa$.

\begin{figure}
    \centering
    \includegraphics[width=0.75\textwidth]{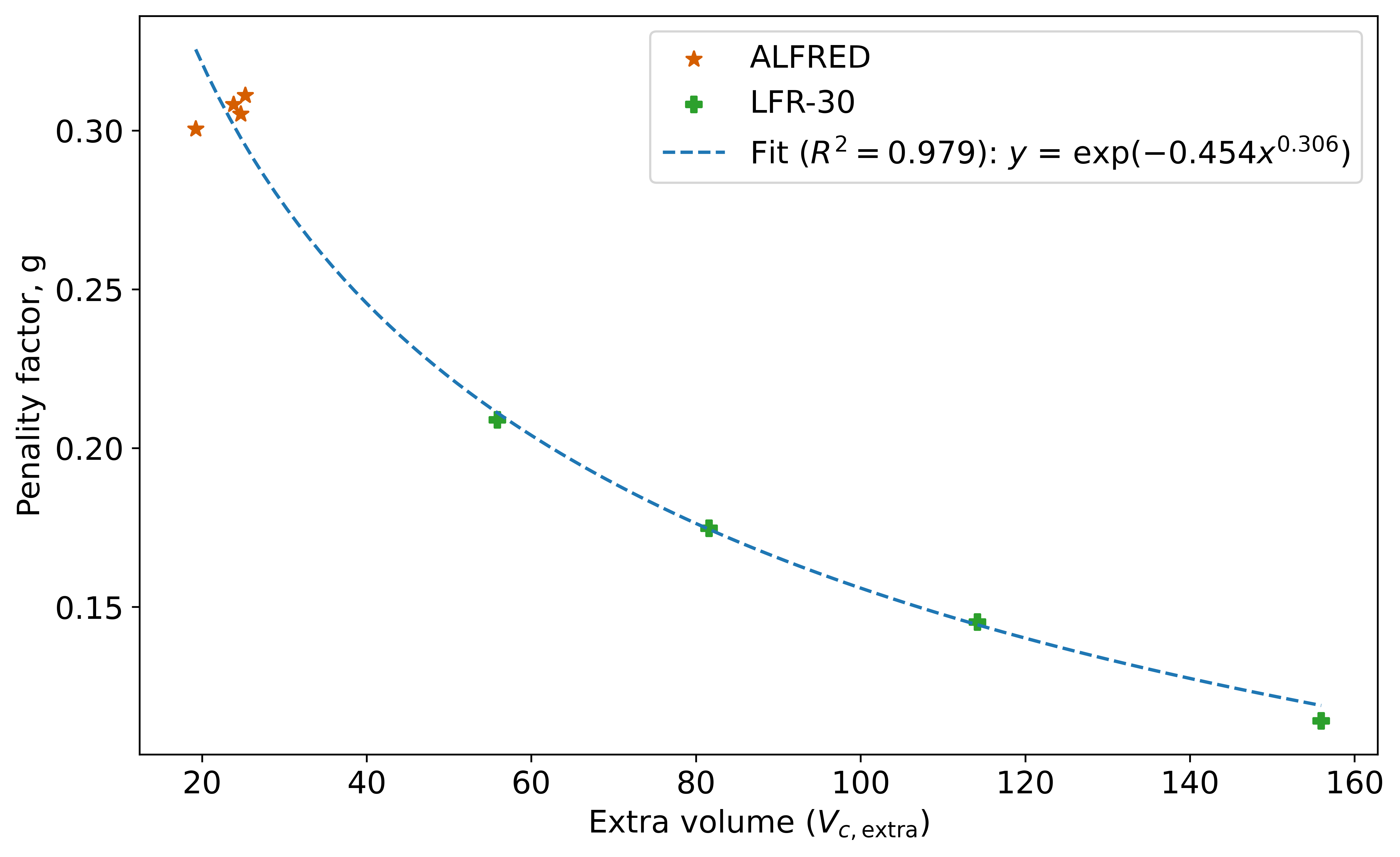}
    \caption{\small Penalty factor, $g$, as a function of the extra volume of lead in the LFR-30 (orange) and ALFRED (blue) assembly configurations.}
    \label{fig:PenalityFactorFit}
\end{figure}

By rearranging Eqn.~\ref{eqn:PhotonRelatedHeating}, the factor $g$ was determined for the LFR-30 and ALFRED configurations. The data was then plotted against $V_{c,\text{extra}}$ for both models and fitted using a decaying exponential, as displayed in Fig.~\ref{fig:PenalityFactorFit}. This process aimed to derive an expression for the penalty factor that can be adapted to different types of fuel assembly.

In the ALFRED model, the extra volume comes from lead bordering the hexagonal wrapper, from both inside and outside of the assembly's wrapper, and from lead between the inner-most circle of fuel pins and the central circular channel. For the LFR-30 model an additional contribution originates from lead within the central channel. Therefore, due to the additional volume of lead in the center, the LFR-30 experiences a lower heating density compared to that in the ALFRED configurations, implying a lower penalty factor. Using values of $g$ from the fit, Tab.~\ref{tab:PhotonRelatedHeatingResults} reports the expected and predicted results for the photon-related coolant heating, presenting good agreement.

\begin{table}[htbp]
\renewcommand*{\arraystretch}{1.4}
\caption{\small Results from the approximation to photon-related coolant heating for the LFR-30 and ALFRED configurations.}
\centering
    \begin{tabular}{cccccc}
    \hline
    & \multicolumn{2}{c}{LFR-30} && \multicolumn{2}{c}{ALFRED} \\
    \cline{2-3}
    \cline{5-6}
    & $\frac{V_{c, \text{extra}, A}}{V_{c,  FC}}$ & \% difference && $\frac{V_{c, \text{extra}, A}}{V_{c,  FC}}$ & \% difference \\
    \hline
    Original        & 96.72  & -0.0825 && 26.13 & -0.4875 \\
    Configuration 1 & 141.21 & 0.0204  && 32.36 & 0.1605 \\
    Configuration 2 & 197.60 & -0.0824 && 34.33 & 0.4591 \\
    Configuration 3 & 269.88 & -0.7576 && 33.56 & 0.2507 \\
    \hline
    \end{tabular}
    \label{tab:PhotonRelatedHeatingResults}
\end{table}

To complete the relation, the next step involved approximating heating in non-coolant materials from the pin cell calculations which encompass the fuel, the cladding made from AIM1 steel and the helium-filled gap. From the results obtained in Sec.~\ref{sec:Edep} and in Tab.~\ref{tab:tab2}, the majority of the energy deposition across these materials occurs in the fuel due to neutron-induced reactions, with fission being the largest contributing channel. The energy deposited in non-coolant material is therefore largely influenced by the number of fuel pins in the assembly and can be approximated to first order as
\begin{equation}
    E_{nc, A} = N_{FC}E_{nc, FC},
    \label{eqn:NoncoolantApprox}
\end{equation}
where $E_{nc}$ is the heating in non-coolant materials. To verify this approximation, the predicted number of fuel pins obtained by dividing Eqn.~\ref{eqn:NoncoolantApprox} by $E_{nc, FC}$ were calculated and compared to the expected number. The outcomes are showcased in Tab.~\ref{tab:NonCoolantHeatingResults} and indicate discrepancies smaller than $1.5\%$, supporting the approximation.
\begin{table}[htbp]
    \renewcommand*{\arraystretch}{1.4}
    \caption{\small Results from the non-coolant heating approximation for the LFR-30 and ALFRED configurations.}
    \centering
    \begin{tabular}[\textwidth]{cccc}
    \hline
    \multicolumn{4}{c}{\textbf{LFR-30}} \\
    \hline
        & \begin{tabular}[c]{@{}c@{}} Number \\ of fuel pins\end{tabular} & \begin{tabular}[c]{@{}c@{}} Predicted \\ number of fuel pins\end{tabular} & \% difference \\
    \hline
    Original        & 234 & 232.8 & 0.5311 \\
    Configuration 1 & 210 & 208.5 & 0.7308 \\
    Configuration 2 & 180 & 178.2 & 1.0021 \\
    Configuration 3 & 140 & 138.1 & 1.3968 \\
    \hline
    \multicolumn{4}{c}{\textbf{ALFRED}} \\
    \hline
        & \begin{tabular}[c]{@{}c@{}} Number \\ of fuel pins\end{tabular} & \begin{tabular}[c]{@{}c@{}} Predicted \\ number of fuel pins\end{tabular} & \% difference \\
    \hline
    Original        & 126 & 125.6 & 0.3325 \\
    Configuration 1 & 120 & 119.5 & 0.4455 \\
    Configuration 2 & 108 & 107.4 & 0.5305 \\
    Configuration 3 & 90  & 89.5 & 0.6103 \\
    \end{tabular}
    \label{tab:NonCoolantHeatingResults}
\end{table}

Combining these relations, the fraction of coolant heating in an assembly can be expressed as:
\begin{equation}
    \zeta_{A} \simeq \frac{E_{c,\gamma,FC}\kappa}{E_{nc,FC} + E_{c,\gamma,FC}\kappa},
\end{equation}
which can be rewritten as a function of the fraction of coolant heating in the fuel cell as
\begin{equation}
    \zeta_{A} = \frac{\zeta_{FC} \kappa}{1 - (1 - \kappa)\zeta_{FC}}.
\end{equation}
The results obtained from the relation are recorded in Tab.~\ref{tab:CorrelationResults} and exhibit differences smaller than $1.6\%$ with the true coolant heating fraction.
\begin{table}[htpb]
    \renewcommand*{\arraystretch}{1.4}
    \caption{\small Relation results demonstrating the successful prediction of fractional coolant heating in the LFR-30 and ALFRED assemblies from pin cell calculations.}
    \centering
    \begin{tabular}[\textwidth]{cccc}
    \hline
    \multicolumn{4}{c}{\textbf{LFR-30}} \\
    \hline
        & \begin{tabular}[c]{@{}c@{}} Coolant heating \\ fraction [\%] \end{tabular} & \begin{tabular}[c]{@{}c@{}} Predicted coolant \\ heating fraction [\%] \end{tabular} & \% difference \\
    \hline
    Original        & 6.161 & 6.125 & 0.5863 \\
    Configuration 1 & 6.349 & 6.284 & 1.0226 \\
    Configuration 2 & 6.602 & 6.501 & 1.5511 \\
    Configuration 3 & 6.970 & 6.871 & 1.4372 \\
    \hline
    \multicolumn{4}{c}{\textbf{ALFRED}} \\
    \hline
        & \begin{tabular}[c]{@{}c@{}} Coolant heating \\ fraction [\%] \end{tabular} & \begin{tabular}[c]{@{}c@{}} Predicted coolant \\ heating fraction [\%] \end{tabular} & \% difference \\
    \hline
    Original        & 5.385 & 5.394 & -0.1613 \\
    Configuration 1 & 5.492 & 5.460 & 0.5826 \\
    Configuration 2 & 5.572 & 5.520 & 0.9517 \\
    Configuration 3 & 5.648 & 5.602 & 0.8216 \\
    \end{tabular}
    \label{tab:CorrelationResults}
\end{table}

\begin{figure}[htpb]
    \centering
    \includegraphics[width=0.8\textwidth]{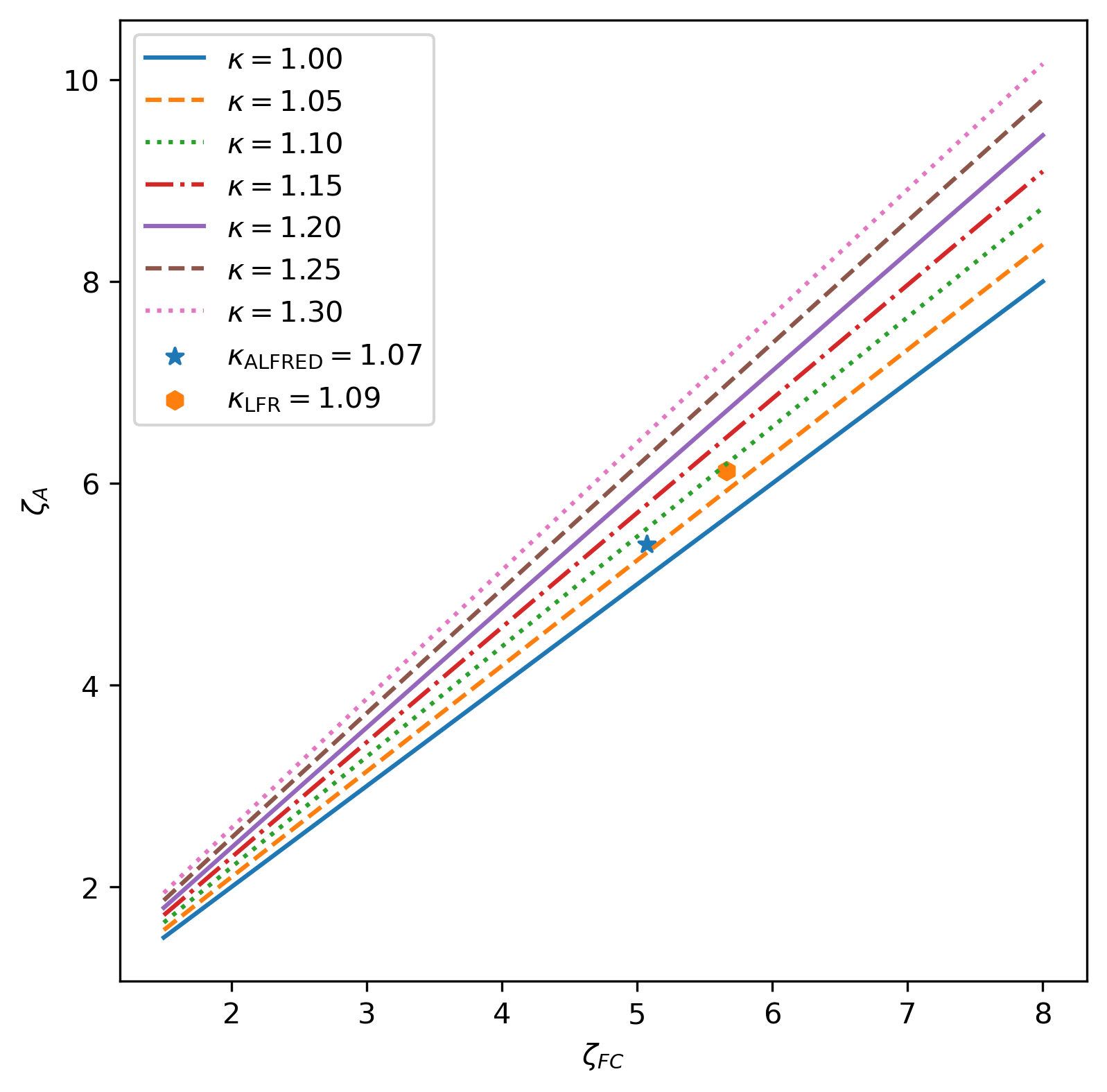}
    \caption{\small Fuel assembly coolant heating as a function of fuel cell coolant heating for different plausible values of $\kappa$. The $\kappa$ values for the original LFR30 and ALFRED fuel assemblies are indicated for reference.}
    \label{fig:KappaCharacterisation}
\end{figure}

Using the relation, the coolant heating in the assembly is plotted in Fig~\ref{fig:KappaCharacterisation} for different values of $\kappa$. Here, $\kappa=1$ corresponds to fuel cells arranged in an infinite lattice, while $\kappa > 1$ reflects different fuel assembly designs. Therefore, the expected coolant heating in a specific fuel assembly model can be estimated for fuel cells of varying coolant heating fractions since each created design will have its corresponding $\kappa$.

\section{Conclusion}
This work aimed at the determination of the fraction of energy deposited in the coolant in a Lead-cooled Fast Reactor and at the understanding of the underlying fundamental physical mechanisms. The studies were conducted using OpenMC with the pin geometry, after assessing its superior performances with respect to the lattice geometry. The calculations estimated the fraction of energy deposited in lead to be $\approx$5.6\%, a value higher than that observed in PWRs~\cite{KINAST2021}. This discrepancy is primarily attributed to the stronger interaction of photons with a high-Z material via the photoelectric effect. Bremsstrahlung was found to give a minor contribution to the energy deposit, while it does influence the photon count and therefore the distribution of total deposited energy among particle species. A parametric study on the influence of coolant temperature on heating revealed no significant impact. Variations in the distance between fuel rods resulted in a change of about 17\% of the fraction of deposited energy. This behaviour stems from a mass effect, where an increase in the quantity of lead results in more interactions within it.
\\
A relation was developed to predict the coolant heating fraction in general fuel assembly designs based on geometrical features. This formula is proposed as a tool for engineering purposes, offering the ability to reduce the computational cost associated with simulating entire fuel assemblies and providing better insights into coolant heating within these assemblies.
\\
In this work, the energy deposited by the decay of unstable isotopes in the fuel, as well as the heat resulting from the activation of the cladding and lead, was neglected, leading to a systematic underestimation in the results. Further studies will focus on estimating these effects and performing burn-up calculations to enhance our understanding of energy deposition in LFRs.

\section*{Acknowledgments}
The authors are thankful to Matteo Zammataro and Matteo Falabino for fruitful discussions and for supporting the preparation of the calculations.

\section*{Data availability statement}
No Data associated in the manuscript.

\bibliographystyle{ieeetr}
\bibliography{bibliography.bib}

\begin{thebibliography}{10}

\bibitem{IEA2022}
{International Energy Agency}, {\em Nuclear Power and Secure Energy
  Transitions}.
\newblock 2022.

\bibitem{PIORO2016}
I.~Pioro, ``{2 - Introduction: Generation IV International Forum},'' in {\em
  Handbook of Generation IV Nuclear Reactors} (I.~L. Pioro, ed.), Woodhead
  Publishing Series in Energy, pp.~37--54, Woodhead Publishing, 2016.

\bibitem{GIF2014}
{GIF}, ``{A Technology Roadmap for Generation IV Nuclear Energy Systems},''
  {Technical Report} NEA-GIF-2014-1, {NEA/OCSE} for the {G}eneration {IV}
  {I}nternational {F}orum, 2014.

\bibitem{SMITH2016}
C.~Smith and L.~Cinotti, ``{6 - Lead-cooled fast reactor},'' in {\em Handbook
  of Generation IV Nuclear Reactors} (I.~L. Pioro, ed.), Woodhead Publishing
  Series in Energy, pp.~119--155, Woodhead Publishing, 2016.

\bibitem{KINAST2021}
S.~Kinast and D.~Tomatis, ``{Energy deposition in coolant of PWR under normal
  operation and accident conditions},'' {\em Nuclear Engineering and Design},
  vol.~384, p.~111479, 2021.

\bibitem{ALEMBERTI2014}
A.~Alemberti, G.~Villabruna, P.~Agostini, G.~Grasso, I.~Turcu, and
  M.~Constantin, ``{ALFRED and the FALCON Consortium},'' in {\em Third
  International Scientific and Technical Conference ``Innovative designs and
  technologies of nuclear power'' (ISTC NIKIET-2014)}, (Moscow, Russia),
  October 7-10 2014.

\bibitem{ROMANO2015}
P.~K. Romano, N.~E. Horelik, B.~R. Herman, A.~G. Nelson, B.~Forget, and
  K.~Smith, ``{OpenMC: A state-of-the-art Monte Carlo code for research and
  development},'' {\em Annals of Nuclear Energy}, vol.~82, pp.~90--97, 2015.
\newblock Joint International Conference on Supercomputing in Nuclear
  Applications and Monte Carlo 2013, SNA + MC 2013. Pluri- and
  Trans-disciplinarity, Towards New Modeling and Numerical Simulation
  Paradigms.

\bibitem{BROWN20181}
D.~Brown {\em et~al.}, ``{ENDF/B-VIII.0: The 8th Major Release of the Nuclear
  Reaction Data Library with CIELO-project Cross Sections, New Standards and
  Thermal Scattering Data},'' {\em Nuclear Data Sheets}, vol.~148, pp.~1--142,
  2018.
\newblock Special Issue on Nuclear Reaction Data.

\bibitem{FORMAGGIO2012}
J.~A. Formaggio and G.~P. Zeller, ``{From eV to EeV: Neutrino cross sections
  across energy scales},'' {\em Rev. Mod. Phys.}, vol.~84, pp.~1307--1341, Sep
  2012.

\bibitem{osti_1338791}
R.~Macfarlane {\em et~al.}, ``{The NJOY Nuclear Data Processing System, Version
  2016},'' 1 2017.

\bibitem{ROMANO2020}
P.~Romano, A.~Johnson, A.~Lund, and J.~Liang, ``{Energy Deposition in the
  OpenMC Monte Carlo Particle Transport Code},'' {\em Transactions}, vol.~123,
  pp.~1345--1348, 2020.

\bibitem{LEPPANEN2015142}
J.~Leppänen, M.~Pusa, T.~Viitanen, V.~Valtavirta, and T.~Kaltiaisenaho, ``The
  {S}erpent {M}onte {C}arlo code: {S}tatus, development and applications in
  2013,'' {\em Annals of {N}uclear {E}nergy}, vol.~82, pp.~142--150, 2015.
\newblock Joint International Conference on Supercomputing in Nuclear
  Applications and Monte Carlo 2013, SNA + MC 2013. Pluri- and
  Trans-disciplinarity, Towards New Modeling and Numerical Simulation
  Paradigms.

\bibitem{jne2020020}
D.~P. Griesheimer, S.~J. Douglass, and M.~H. Stedry, ``Self-{C}onsistent
  {E}nergy {N}ormalization for {Q}uasistatic {R}eactor {C}alculations,'' {\em
  Journal of Nuclear Engineering}, vol.~2, no.~2, pp.~215--224, 2021.

\bibitem{LUND2018}
A.~Lund and P.~Romano, ``{Implementation and Validation of Photon Transport in
  OpenMC},'' Tech. Rep. ANL/MCS-TM-381, Argonne National Laboratory, Dec 2018.

\bibitem{CINOTTI2010}
L.~Cinotti, C.~F. Smith, C.~Artioli, G.~Grasso, and G.~Corsini, ``{Lead-Cooled
  Fast Reactor (LFR) Design: Safety, Neutronics, Thermal Hydraulics, Structural
  Mechanics, Fuel, Core, and Plant Design},'' in {\em Handbook of Nuclear
  Engineering} (D.~G. Cacuci, ed.), pp.~2749--2840, Boston, MA: Springer US,
  2010.

\bibitem{GODFREY14}
A.~Godfrey, ``{VERA core physics benchmark progression problem
  specifications},'' {Technical Report} CASL-U-2012-0131, {Consortium for
  Advanced Simulation of LWRs(CASL), Rev. 4}, 2014.

\bibitem{HAGHIGHAT2020}
A.~Haghighat, {\em {Monte Carlo Methods for Particle Transport}}.
\newblock Taylor \& Francis, 2020.

\bibitem{sartori1990standard}
E.~Sartori, ``{Standard energy group structures of cross section libraries for
  reactor shielding, reactor cell and fusion neutronics applications:
  VITAMIN-J, ECCO-33, ECCO-2000 and XMAS},'' {\em {JEF/DOC-315, Revision}},
  vol.~3, 1990.

\bibitem{CASTAGNA}
C.~Castagna, D.~Tomatis, S.~Kinast, and E.~Gilad, ``{Analysis of the
  correlation for energy deposition in PWR coolant by coupled neutron-photon
  transport calculations},'' {\em {Progress in Nuclear Energy}}, vol.~172,
  p.~105173, 2024.

\end{thebibliography}

\end{document}